\pdfoutput=1

\documentclass[prodmode,acmtochi]{acmsmall} 

\usepackage[ruled]{algorithm2e}

\SetAlFnt{\small}
\SetAlCapFnt{\small}
\SetAlCapNameFnt{\small}
\SetAlCapHSkip{0pt}
\IncMargin{-\parindent}

\acmYear{2017} \acmVolume{1} \acmNumber{1} \acmArticle{1} \acmMonth{1} 

\usepackage[ruled]{algorithm2e}
\usepackage{booktabs}
\usepackage{tabu}
\usepackage{color}
\usepackage[dvipsnames]{xcolor}
\usepackage{multirow}
\usepackage{colortbl}
\usepackage[most]{tcolorbox}
\usepackage{enumitem}
\usepackage{pifont}

\SetAlFnt{\small}
\SetAlCapFnt{\small}
\SetAlCapNameFnt{\small}
\SetAlCapHSkip{0pt}
\IncMargin{-\parindent}

\definecolor{lightgray}{gray}{0.95}
\long\def\ignore#1{}
\newcommand{\nam}[1]{{\color{black}{#1}}}
\newcommand{\zoya}[1]{{\color{black}{#1}}}
\newcommand{\rev}[1]{{\color{black}{#1}}}
\newtcolorbox{myframe}[1][]{
  enhanced,
  arc=0pt,
  outer arc=0pt,
  colback=lightgray,
  boxrule=0.2pt,
  #1
}

\definecolor{Gray}{gray}{0.9}

\setcopyright{acmcopyright}

\doi{10.1145/3131275}

\issn{1234-56789}

\begin{document}

\title{BubbleView: an interface for crowdsourcing image importance maps and tracking visual attention}
\author{
Nam Wook Kim*
\affil{Harvard SEAS}
Zoya Bylinskii*
\affil{MIT CSAIL}
Michelle A. Borkin
\affil{Northeastern CCIS}
Krzysztof Z. Gajos
\affil{Harvard SEAS}
Aude Oliva
\affil{MIT CSAIL}
Fredo Durand
\affil{MIT CSAIL}
Hanspeter Pfister
\affil{Harvard SEAS}
}

\markboth{Kim \& Bylinskii,  et al.}{BubbleView}

\begin{abstract}
In this paper, we present BubbleView, \rev{an alternative methodology for eye tracking using discrete mouse clicks to measure which information people consciously choose to examine. BubbleView is a mouse-contingent, moving-window interface in which} participants are presented with a series of blurred images and click to reveal ``bubbles'' - small, circular areas of the image at original resolution, similar to having a confined area of focus like the eye fovea. 
\rev{Across 10 experiments with 28 different parameter combinations}, we evaluated BubbleView on a variety of image types: information visualizations, natural images, static webpages, and graphic designs, and compared the clicks to eye fixations collected with eye-trackers in controlled lab settings. We found that BubbleView clicks can both (i) successfully approximate eye fixations on different images, and (ii) be used to rank image and design elements by importance. BubbleView is designed to \rev{collect clicks on static images}, and works best for defined tasks such as describing the content of an information visualization or measuring image importance. \nam{BubbleView data is cleaner and more consistent than related methodologies that use continuous mouse movements.} 
\zoya{Our analyses validate the use of mouse-contingent, moving-window methodologies as approximating eye fixations for different image and task types.}
\end{abstract}

%
%
%


%
%



\begin{bottomstuff}
*Equal contribution.

\end{bottomstuff}

\maketitle

\section{Introduction}
\label{sec:introduction}

\begin{figure}[h]
 \centering
\includegraphics[width=1\textwidth]{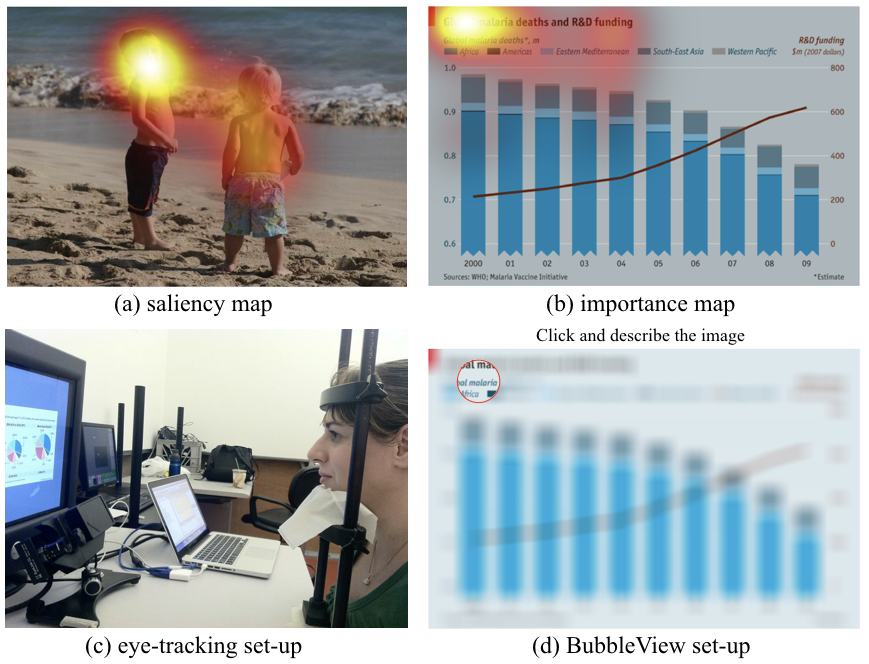}
 \caption{ Just as the pattern of human eye fixations can be used as a heatmap of saliency for an image (a), the pattern of BubbleView clicks can be used as a heatmap of importance for an image (b). An eye tracking set-up \rev{(pictured: EyeLink1000)} is a way to collect human eye fixations \rev{in the lab setting} (c), whereas the BubbleView interface can be launched online and feasibly scale up the collection of crowdsourced data (d).}
 \label{fig:overview}
\end{figure}

Eye tracking is a technique to measure an individual's eye movements, visual attention, and focus. This experimental methodology has proven useful for studying the cognitive processes involved in visual information processing, including which visual elements people look at first and spend the most time on~\cite{jacob2003eye,majaranta2014eye}. Eye tracking is widely used for conducting usability studies for human-computer interfaces~\cite{jacob2003eye,nielsen2010eyetracking}, for designing gaze-based and attention-aware user interfaces~\cite{majaranta2014eye,lutteroth2015gaze} \rev{or for collecting gaze data to build saliency prediction models~\cite{borji2015cat2000,judd2012benchmark}}.

\rev{
Commercial eye-trackers mostly use specialized hardware such as advanced infrared sensors and high-quality cameras to accurately track eye positions and movements~\cite{al2013eye}. However, they often require high-cost equipment and invasive calibrations (e.g., EyeLink, ISCAN), which means it is difficult to scale to large scale studies beyond controlled lab environments. Recent appearance-based methods attempt to address this issue by enabling eye tracking on affordable cameras built into personal devices~\cite{xu2015turkergaze,huang2015tabletgaze,krafka2016eye}. However, these methods have not yet seen widespread adoption, as they still suffer in accuracy and robustness, and impose set-up constraints (camera quality, lighting conditions). 

On the other hand, cursor-based attention tracking is based on the correlation between gazes and cursor locations~\cite{huang2012user,rodden2008eye,guo2010towards} and reduce the need to handle variations in real-world settings in camera-based methods; e.g., calibrations, ambient lighting, etc. The most popular cursor-based approach uses a moving window continuously following the position of the cursor to reveal a portion of the screen in normal resolution~\cite{jansen2003tool}. 
}


Our \textbf{BubbleView} methodology is a \zoya{cursor-based, moving-window} approach to collect clicks on static images as a proxy for eye fixations. 
BubbleView presents blurred images and allows participants to click around to reveal small circular ``bubble'' regions of the image at the original resolution (Figure~\ref{fig:overview}). This is intended to loosely approximate a blurred periphery and the confined area of focus of the human eye fovea. 





\rev{Compared to natural viewing, BubbleView and related cursor-based methodologies slow down the exploration patterns of participants, because choosing where to move the mouse and click is a slower cognitive process than moving eyes around an image. Because of this, we refer to the pattern of BubbleView clicks on an image as the \textbf{importance map} for the image. 
We intend for importance to encapsulate image regions that are not only more attention grabbing initially (salient), but also regions that people spend more time on because they are more relevant, or interesting, to the task-at-hand.}

BubbleView is especially well suited to capturing image regions of most importance \rev{when a directed task is provided (as compared to free viewing)}. Our initial target setting, first presented in~\citeN{kim2015crowdsourced} was to show that BubbleView clicks can provide a good approximation for eye movements when participants are asked to describe the content of information visualizations (graphs, charts, tables). In~\citeN{beyondmem}, we further showed that knowing where people look can provide clues about what they store in memory and recall about an information visualization. Like eye \nam{tracking}, BubbleView can provide important insights about human perception and cognition, but at a lower data collection cost than eye tracking. It can easily scale up data collection to many participants and images, and be launched remotely to enable online crowdsourcing.


In this paper, we validate that BubbleView generalizes to approximating eye fixations on different image types and under different task constraints. 
Specifically, we show that:
\begin{itemize}[label=\ding{212}]
\item BubbleView clicks can successfully approximate eye fixations on information visualizations, natural images, and websites, in both a free-viewing condition and with a description task;
\item Compared to related methodologies based on a moving-window approach~\cite{jiang2015salicon}, BubbleView clicks provide more reliable and less noisy data;
\item The number of BubbleView clicks in different image regions can be used to measure the relative importance of those image regions.
\end{itemize}

We present the BubbleView methodology with the interested experimenter in mind who may consider it for crowdsourcing an experiment\rev{, or for an evaluation that would typically be conducted with an eye tracker in a conventional laboratory setting}. 
\zoya{While prior work contains some initial validation that a cursor-based interface can serve as a proxy for eye tracking~\cite{bednarik2007validating,jiang2015salicon,kim2015crowdsourced}, we conducted an extensive quantitative analysis by running 10 experiments with 28 different parameter combinations, on Amazon's Mechanical Turk.}
Our experiments \nam{were} carried out on 5 different datasets, spanning information visualizations~\cite{beyondmem}, natural images~\cite{xu2014predicting,jiang2015salicon}, static webpages~\cite{shen2014webpage}, and graphic designs~\cite{odonovan}. We varied task type (free-viewing, describing) and task \nam{duration}, image blur kernel, and bubble radius. We compared BubbleView clicks not only to eye fixations~\cite{beyondmem,xu2014predicting}, but also to mouse movements~\cite{jiang2015salicon}, and to explicit importance annotations~\cite{odonovan}.  
Our contributions include: 
\begin{enumerate}
\item The BubbleView interface which can be launched online for the cheap, feasible collection of crowdsourced data, provided at \underline{massvis.mit.edu/bubbleview};
\item A thorough analysis of how different experimental parameters affect BubbleView click data, and guidelines about how to choose an appropriate setting of parameters for a given experiment;
\item A discussion of how BubbleView can be used \rev{to approximate} eye fixations collected in a controlled lab setting;
\item A proposed list of applications of the BubbleView methodology, including for the measurement of image importance, image-based question-answering tasks, and training computational models of saliency/importance.
\end{enumerate}




\section{Related Work}
\label{sec:related-work}

\zoya{The original idea of ``bubbles'' comes from the work of \citeN{GossSchyn} who displayed masked images, punctured by randomly-located Gaussian windows (termed bubbles), and measured participant performances on categorization tasks to determine image regions important to the tasks. \citeN{deng2013fine} modified this methodology to allow participants to control the location of bubbles on blurred images, and reveal image regions in order to complete a fine-grained object recognition task. We further extended the bubbles technique by having participants click to expose image regions in order to describe information visualizations~\cite{kim2015crowdsourced}. In this paper we evaluate the BubbleView methodology with both description and free-viewing tasks on information visualizations, natural images, and graphic designs, to validate that it can be used for discovering relevant image regions. In these settings, BubbleView is similar to other cursor-based, moving-window approaches which expose image regions depending on user-defined cursor positions~\cite{jansen2003tool,schulte2011flashlight}. Because we vary the resolution of the image depending on where the cursor is located, BubbleView can also be classified as a mouse-contingent, multiresolutional display (similar to~\citeN{jiang2015salicon}, which we compare to in this paper). In this section we review other gaze tracking techniques, including eye tracking, cursor-based and appearance-based approaches. We also discuss the relationship of gaze tracking and mouse tracking to saliency.}

\subsection{Eye movements and cognitive tasks}
A significant amount of research has been conducted on the connection between eye movements and various cognitive tasks: the eyes can provide important clues about how visual perception proceeds as a human looks at images~\cite{holmqvist2011eye,just1976eye,hayhoe2004advances,kowler1989role,noton1971scanpaths}. This area of research is so established and diverse that we refer the reader to some representative papers reporting on the utility of eye movements for studying human perception and cognition in the context of user interfaces~\cite{bergstrom2014eye,duchowski2002breadth,goldberg1999computer,graf1989ergonomic,poole2006eye,jacob2003eye,bruneau2002eyes,rensink2011management}, web search\rev{~\cite{cutrell2007you,goldberg2002eye}}, web browsing~\cite{cowen2002eye,josephson2002visual,pan2004determinants}, problem solving~\cite{grant2003eye}, reading~\cite{rayner1998eye}, advertisements~\cite{rayner2001integrating}, and visualizations~\cite{beyondmem,etvisChapter,pohl2009,Kim2012,Huang2007}. These papers show that aside from providing information about how human perception proceeds, eye movements can also provide insights about the effectiveness of different visual content, or the usability of interfaces. Because of all the potential use cases, researchers have also sought ways to more efficiently collect eye movements without having to rely on standard eye tracking.\\

\rev{\subsection{Cursor-based attention tracking}}
There has been a significant effort to find cheap, nonintrusive, and more scalable alternatives to collect human attentional data. \rev{Cursor-based techniques are a particularly suitable alternative for scaling to large web-based studies.}

The \textbf{moving-window} approach is a popular cursor-based technique in which a limited amount of information is visible through a variable size window \nam{continuously following a cursor position}~\cite{mcconkie1975span,rayner2014gaze}. 
Inspired by the moving-window model, \citeN{jansen2003tool} developed a computer program called Restricted Focus Viewer (RFV) that takes an image, blurs it, and reveals only a restricted \rev{block} of the image, allowing a user to move the region using a mouse~\cite{jansen2003tool,tarasewich2005enhanced,bednarik2007validating,blackwell2000restricted}. Commercial software for tracking user attention has also built on the same idea \rev{(e.g., Attensee\footnote{\rev{Attensee (\url{http://www.attensee.com}) is a commercial solution based on the idea of Flashlight~\cite{schulte2011flashlight}, an open-source research tool: \url{https://github.com/michaelschulte/flashlight}.}})}. The mouse-contingent methodology has been employed to investigate cognitive behaviors of
users in diverse contexts such as diagrammatic reasoning and program debugging, and to study the usability of web sites~\cite{jansen2003tool,bednarik2005effects,tarasewich2005enhanced}. 

Recent studies have made further improvements. \rev{SALICON \cite{jiang2015salicon}} implemented moving-window, multi-resolution blur on images to attempt to simulate the fall-off in acuity of peripheral vision. 
On the other hand, \citeN{lagun2011viewser} directly preprocessed web search results to show one result and blur the other results based on a user's viewport; however, this method is not intended to approximate the human fovea as it shows an entire DOM element at a time.  All these recent studies were conducted online with hundreds to thousands of participants, proving the scalability of their methods.

\rev{There is also a rich history of work in the space of gaze-contingent multiresolutional displays, where the moving-window approach is guided by gaze. We refer the reader to a review by~\citeN{reingold2003gaze}. These approaches complement, rather than replace, standard eye-tracking techniques, and have different motivations: bandwidth and processing savings. However, this line of work contains a related investigation of multiresolutional blur to approximate the peripheral visual system. Whether cursor-based or gaze-based, a moving-window approach slows down visual exploration patterns relative to natural viewing and can be used to discover the most important or relevant image regions.}

Aside from the moving-window model for image exploration, other works also investigated the relationship between cursor movements and gaze positions, mostly focusing on web browsing~\cite{chen2001can} and search tasks~\cite{rodden2008eye,guo2010towards,huang2011no,huang2012user}. 
\citeN{chen2001can} found a high correlation between cursor and gaze locations. \citeN{rodden2008eye} found that cursor and gaze are better aligned along the vertical dimension, while \citeN{guo2010towards} also found a similar result in their study of predicting eye-mouse coordination. \citeN{huang2012user} found that people's cursors lag behind their gazes and there are individual differences in the distance between the cursor and gaze positions. 

\nam{These cursor-based techniques have succeeded in providing an affordable and scalable alternative to eye tracking, but prior work has two key limitations. First, a moving window approach requires complicated post-processing of mouse movement data to extract mouse positions (e.g., SALICON). Second, evaluations of existing techniques have been mostly limited to simple aggregate comparisons with ground-truth eye tracking data on a specific set of images (i.e., natural images or webpages) with a fixed setting of parameters (e.g., blur kernel, bubble size).

With BubbleView, we overcome the first limitation by collecting discrete clicks instead of continuous mouse trajectories. This enables a more explicit record of points of interest without the need for post-processing noisy mouse movement data. To address the second limitation, we also systematically evaluate the effect of different parameters and task settings on the ability of a cursor-based methodology to approximate eye movements.}
We compare our methodology to eye fixations on a diverse set of image stimuli with different parameters to find the best settings under different task conditions. 
Our findings are likely to generalize to other related mouse-contingent displays.

\rev{\subsection{Appearance-based gaze tracking}

Another line of work has been devoted to non-intrusive, appearance-based gaze estimation, where images of the eyes are post-processed using computer vision techniques to determine gaze location. \zoya{This type of gaze estimation often involves collecting a training dataset with a standard eye-tracker, training a computer vision model to map eye images to gaze coordinates, and using this model at test-time to directly infer gaze positions from a video stream of the eyes (e.g., captured via a webcam). At test time, these approaches do not require specialized eye tracking hardware (i.e., high quality special cameras, infrared sensors, and head mounting devices) and allow users to move their heads freely.} 

Early gaze tracking models were mostly based on relatively small training datasets collected through lab studies. For example, \citeN{baluja1994non} collected 2000 images of the eyes for four postures by instructing a participant to visually track a moving cursor and built a neural network model to estimate gaze locations. Recent methods attempt to build gaze tracking models on large datasets to improve accuracy as well as to work in real-world settings. 
\citeN{mora2014eyediap} constructed a database to enable comparison across different gaze tracking algorithms for variations including head poses, individual differences, and ambient and sensing conditions. \citeN{zhang2015appearance} developed an appearance-based gaze estimation method using multimodal convolutional neural networks. Their model was trained on a hundred thousand images from 15 laptop users for several months using built-in cameras in laptops, accounting for realistic variability in illumination and appearance. \citeN{huang2015tabletgaze} similarly built a large gaze dataset and a gaze tracking algorithm for tablet users. While the two studies are still limited to datasets collected through labs, other works leverage online crowdsourcing to further extend the scale of gaze datasets. \citeN{xu2015turkergaze} developed a webcam-based eye tracking game running in a browser on a remote computer. Their crowdsourced experiments could collect gaze data cheaper and faster than lab studies. \citeN{papoutsaki2016webgazer} also designed a similar webcam-based eye tracking system. \citeN{krafka2016eye} collected eye tracking on over 2.5M frames using a mobile application and online, and developed a gaze prediction algorithm based on convolutional neural networks, while achieving
state-of-the-art results.
}


\rev{All of the above approaches have yet to reach the level of tracking accuracy and robustness possible with dedicated eye tracking hardware.} These approaches also depend on either some initial calibration or have constraints on a participants' set-up: network connection, camera quality, and restricted range of face location relative to screen. \rev{As a result, we have not yet seen widespread adoption of appearance-based gaze tracking.}
Additionally, the camera-based gaze tracking approaches have the downside of requiring the capture of participants' face images throughout the study, which comes with privacy concerns~\cite{liebling2014privacy}.


\subsection{Saliency models and eye tracking datasets}\label{sec:saliency}

In addition to alternative techniques for eye tracking which require human participants, significant progress has been made building computational saliency models to predict eye fixations. Many saliency models are motivated by psychological and neurobiological theories, and make use of both low-level image features (e.g., intensity, color, and orientation) and high-level semantic features (e.g., scenes, objects, and tasks) to approximate the human visual system~\cite{borji2013state,frintrop2010computational}. The performance of these models is usually evaluated against ground-truth eye fixations~\cite{salMetrics_Bylinskii,mit-saliency-benchmark,judd2012benchmark}. 

Models have typically been trained directly on  fixation data collected from eye tracking experiments~\cite{judd2009learning,kienzle2006nonparametric}. \nam{However, good models require large quantities of data, larger than what is practical to collect using conventional eye tracking techniques. To overcome this challenge, a large dataset of mouse movements on natural images was recently released for simulating the natural viewing behavior and subsequently} training computational saliency models. This dataset, dubbed SALICON, was collected using a moving-window methodology~\cite{jiang2015salicon}. Since then, many neural network models of saliency trained on this data~\cite{jiang2015salicon,kruthiventi2015deepfix,pan2016shallow} have achieved state-of-the-art performances on standard saliency benchmarks~\cite{mit-saliency-benchmark}. \rev{\citeN{tavakoli2017saliency} have recently shown that saliency models trained on mouse movements can generalize well to predicting eye fixations.}

While most saliency models are focused on predicting eye fixations on natural scenes, there are relatively few studies that have looked at other image types including webpages, graphic designs, and information visualizations. These images are different from natural images in that they usually contain rich semantic data (e.g., texts, charts, and logos) or different viewing patterns such as top-left bias~\cite{buscher2009you} and banner blindness~\cite{grier2007users}.~\citeN{shen2014webpage} developed a webpage saliency model based on the FiWI dataset, and then improved the model with high-level semantic features (e.g., positional bias and object detectors)~\cite{shen2015predicting}.~\citeN{odonovan} developed a semi-automatic model of importance prediction for graphic designs by training on a crowdsourced dataset of importance annotations. The GDI dataset was collected by asking workers to annotate regions of importance on images using binary masks.~\citeN{xu2016spatio} presented a computational model for predicting visual attention in user interfaces with user interactions. 

We draw on several existing datasets with accompanying attention data and look at how well BubbleView clicks can approximate fixations, mouse movements, and explicit importance annotations. We used the FiWI dataset~\cite{shen2014webpage} (static webpages), OSIE dataset~\cite{xu2014predicting} (natural scenes), and the MASSVIS dataset~\cite{beyondmem} (information visualizations) to evaluate the degree to which BubbleView clicks can approximate eye fixations. We used the SALICON dataset~\cite{jiang2015salicon} to compare our methodology against the moving-window approach. We also used the GDI dataset~\cite{odonovan} (graphic designs) to see whether BubbleView clicks can be used to rank design elements by importance.

\section{BubbleView methodology}
\label{sec:methodology}

BubbleView is an experimental methodology for collecting mouse clicks on images as an \rev{approximation to eye fixations}. 
We first provide some background on human eye movements and perception, before discussing how BubbleView was designed to approximate eye tracking, and how it can be used for running perception experiments.


\subsection{Background: human eye movements and perception}
The human eye consists of light receptor cells that are differently distributed throughout the eye. The clearest and most detailed vision is in the central, \textbf{foveal area}, of the visual field, and blurrier vision is in the larger part of the visual field, which is called the \textbf{peripheral area}. The foveal area captures about 1-2 degrees of visual angle which constitutes less than 8\% of the visual field, but makes up 50\% of the visual information sent to the brain \cite{tobii}. When we move our eyes, we place the foveal region of the eye on different regions of the visual field, bringing them into focus. 

\textbf{Visual angles} are units for measuring the projection of the visual field, as images, on our retina. For a given experimental viewing setup, visual angles can be computed by taking into account the distance to the screen, size and resolution of the image on the screen\footnote{\url{https://github.com/cvzoya/saliency/tree/master/computeVisualAngle}}. The error of professional-grade eye trackers (e.g., EyeLink) is also measured in degrees of visual angle, and is commonly less than 1 degree. 

The pauses in eye movements are called \textbf{fixations}, and the transitions between successive fixations are called \textbf{saccades}. In this paper, we focus on fixations, since they give us the points of interest that the eye has stopped on to bring them into focus. The temporal sequence of fixations\rev{, fixation duration, saccade length, and other features of eye movements} carry a lot of additional information about human perception\rev{~\cite{holmqvist2011eye,just1976eye,jacob2003eye,etvisChapter}} but are beyond the scope of the present work. \rev{We concentrated on the location of fixations, which are most straightforward to analyze~\cite{bruneau2002eyes,jacob2003eye,tobii} and to model computationally~\cite{salMetrics_Bylinskii}.}


\subsection{Designing experiments with BubbleView}

\begin{figure}[h]
 \centering
\includegraphics[width=1\textwidth]{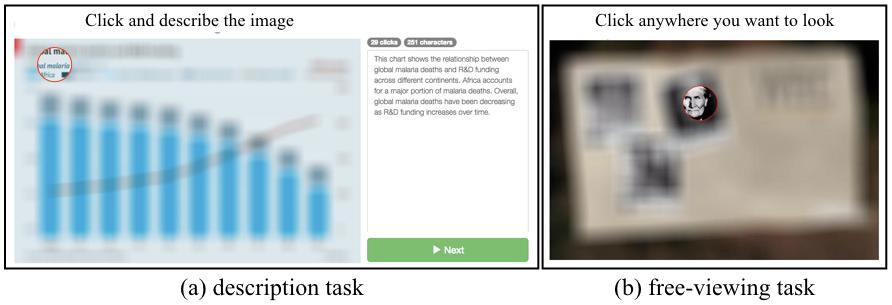}
 \caption{Two different versions of the BubbleView interface for two task types, for gathering task-based (a) and task-free (b) clicks, as approximations to similar eye tracking experiments.}
 \label{fig:interface}
\end{figure}

The BubbleView methodology is intended to approximate a blurred periphery, and users click on images to reveal small, circular regions (``bubbles'') at the original resolution (Figure~\ref{fig:interface}). 
This is similar to having a confined area of focus like the eye fovea.
Different blur levels and bubble sizes can be used to approximate different eye tracking setups, with different visual angles (Figure~\ref{fig:experiment5-setting}). 


In comparison to the moving-window approach which records continuous mouse movements, our approach records discrete mouse clicks where each click represents a conscious choice made by the user to reveal a portion of the image. As the clicks correspond to individual points of interest, we directly compare them to eye fixations. \\

\rev{
\noindent \textit{Tasks and image types for attention experiments} \\
We evaluated our BubbleView interface on two tasks: free-viewing and description, and four image types: natural scenes, information visualizations, static webpages, and graphic designs. Here we discuss the motivations behind these design choices. 

During \textbf{free-viewing}, participants are not given a task but are instructed to freely look around the image. Free-viewing is commonly used in eye tracking experiments to study the human perception of natural scenes, because it can avoid large task-dependent effects. It is often assumed the eyes are drawn to conspicuous image elements, and attention proceeds in a bottom-up manner, guided by the image features rather than a high-level task\footnote{Alternative views posit that free-viewing is not task-free, but permits participants to choose their own internal agendas/tasks~\cite{parkhurst2002modeling,tatler2005visual,tatler2011eye}. Even under this interpretation, averaging data over many participants, each of which may have their own agenda, has the effect of averaging out the task and providing an approximately task-independent aggregate measurement.}. This assumption has motivated the use of free-viewing for collecting ground truth data for saliency datasets~\cite{koehler2014saliency}, where the pattern of eye fixations can be interpreted as the \textbf{saliency map} for the image (Figure~\ref{fig:overview}).
Most saliency datasets have been collected using free-viewing\footnote{A list of eye tracking datasets and their attributes is available at: \url{http://saliency.mit.edu/datasets.html}}. Computational models are in turn trained and tested on saliency datasets as a proxy for human attention. We are similarly motivated by the computational applications that can be built by training models on large attention datasets (e.g.,~\cite{predimportance}).

Compared to natural viewing, cursor-based moving-window methodologies naturally slow down visual exploration patterns. By providing a cognitively-demanding task, these exploration patterns can be slowed down further to bring more intentionality to each click. 
In the \textbf{description} task, participants are required to type a description of the image while using the BubbleView interface to explore the image. The descriptions naturally depend on the image regions clicked on. This task is well suited to images with an underlying message or concept that needs careful examination to decipher. We used the description task with visualization images from the MASSVIS~\cite{beyondmem} dataset\footnote{In the MASSVIS eye-tracking set-up participants also provided image descriptions, but they did so at the end, not during, the viewing session. This is because memorability was part of the original study, whereas it is not here.}, and website images from the FiWI~\cite{shen2014webpage} dataset. We also tested the free-viewing task with the FiWI images, because the eye-tracking data from this dataset was collected with free-viewing, and we wanted to approximate the original experiment. For the same reason, we ran the free-viewing task on natural images from the OSIE~\cite{xu2014predicting} dataset. For the graphic designs in the GDI dataset~\cite{odonovan}, which have importance annotations rather than eye fixations, we chose a free-viewing task. We chose this task because we found that the graphic designs could not be easily summarized by a description (i.e., some images required further context, not all were English, some had few visual elements, etc.). 

Tasks deviating from description and free-viewing are beyond the scope of this paper, although they are common in user interface research~\cite{bergstrom2014eye,jacob2003eye,cutrell2007you,goldberg2002eye}. For instance, for testing websites or application interfaces, participants may be asked to perform tasks such as searching for a particular element or option, navigating to a particular region of the image or page, or answering questions. Related moving-window methodologies have previously been validated in the context of web navigation, program debugging, and question-answering~\cite{bednarik2007validating,jansen2003tool,lagun2011viewser,schulte2011flashlight,tarasewich2005enhanced}. These tasks can be quite specific to the interface being evaluated.
We used two task types that can generalize (without modification) to a large collection of different image types. Our BubbleView tool is available to the research community so future work can investigate the generalizability of this tool for other tasks. \\   
}

\rev{\subsection{Implementation}}

We implemented a web-based BubbleView interface that takes a directory of images as input and 
displays a subset of the images in random sequence, blurring each one. Participants receive a set of task instructions and can click to reveal bubble regions (Figure~\ref{fig:interface}). A demo is available at \underline{massvis.mit.edu/bubbleview}.\\
The experimenter has a choice of parameters:
\begin{itemize}[label=\ding{212}]
\item \textbf{Task type:} the instructions given to participants. We used two different versions of the interface for a description task with an input text field (Figure~\ref{fig:interface}a), and a free-viewing task with no additional inputs from participants (Figure~\ref{fig:interface}b). Alternative tasks are possible.
\item \textbf{Time:} the viewing time per image, which depends on the task. For the description task, we did not constrain the time. For the free-viewing task, we fixed time per image to be either 10 or 30 seconds, depending on the experiment.
\item \textbf{Blur sigma:} the size of the Gaussian blur kernel (in pixels) to apply to each image to mimic peripheral vision. This is a fixed quantity over the whole image, and is constant across all images in the sequence. In our studies, we manually selected a blur value per image dataset to distort image text beyond recognition. We wanted the level of detail to be sufficient for reading only within regions of focus.
\item \textbf{Bubble radius:} the size of the focus area (in pixels) that is deblurred during a click to mimic foveal vision. In our studies, we varied this size depending on other task constraints, but often stayed within 1-2 degrees of visual angle of the eye tracking setups used for the ground-truth eye movement datasets.
\item \rev{\textbf{Mouse modality:} although we originally designed BubbleView for collecting mouse clicks, we extended it to allow bubble regions to be exposed during continuous mouse movements (as in \citeN{jiang2015salicon}). We discuss the differences between the two modalities in Section~\ref{sec:discussion}.}
\end{itemize}

The experimenter may also choose the number of images displayed in a sequence. In our description task, participants were able to continue to the next image after writing a minimum number of characters (150 in our experiments). In the free-viewing task, once the fixed time per image elapsed, the next image in the sequence was presented.

\begin{figure}[h]
 \centering
\includegraphics[width=\textwidth]{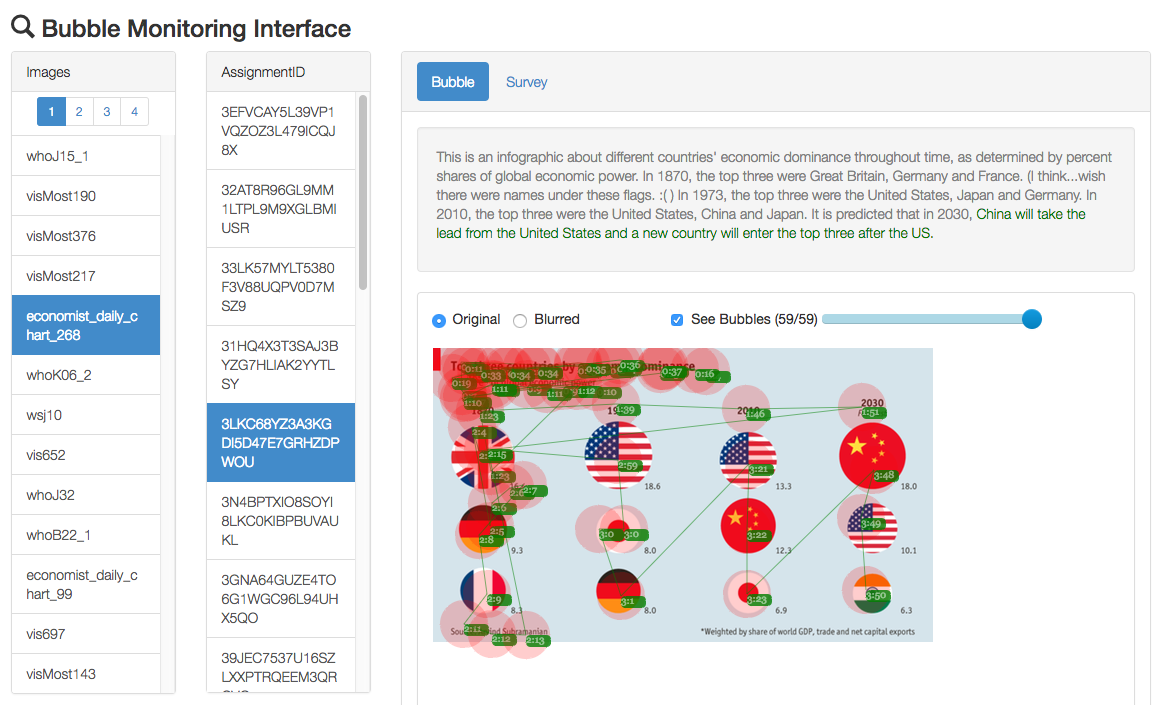}
 \caption{Monitoring interface for manually inspecting the results of experiments. An experimenter can use a slider to explore the temporal sequence and evolution of bubble clicks and description text, for each image and participant.}
 \label{fig:monitoring}
\end{figure}

We also developed a monitoring interface to inspect experimental results (Figure~\ref{fig:monitoring}). The purpose of the interface is to take a quick glance at the bubbles collected, before the main analysis. For each image, the experimenter can see the bubbles and (if applicable) text descriptions generated by each participant. Adjusting the slider allows exploration of the temporal sequence and evolution of bubble clicks and description text over time. The experimenter can also see how the blurred image looked to the participant to investigate why a region may have been clicked. This interface can be used to check if an experiment is running as intended in real time.

\rev{\section{Experiments and Analysis Overview}}
\label{sec:exp_details}

We first presented the BubbleView methodology as a way to approximate eye fixations on information visualizations under a description task~\cite{kim2015crowdsourced}. The present paper is an extension that more systematically explores the BubbleView methodology and measures how varying parameters such as bubble size, image blur, and task timing affects the resulting clicks, and how the number of participants affects the quality of the resulting data. We wanted to test BubbleView for generalizability (i.e., on other image and task types) to see if eye fixations could be well approximated under different settings. 

For our experiments, we used \textbf{Amazon's Mechanical Turk (MTurk)}, an online crowdsourcing platform that makes it easy for experimenters to collect data from as many participants as desired. \textbf{Human Intelligence Tasks (HITs)} are first posted by experimenters. Then participants (MTurk workers) complete the HITs, the results are saved, and payments are issued. Participants remain anonymous to the experimenters.

We compared BubbleView clicks to eye fixations on information visualizations~\cite{beyondmem}, natural scenes~\cite{xu2014predicting}, and webpages~\cite{shen2014webpage}. We also analyzed the relationship between BubbleView and related crowdsourcing methodologies: explicit importance annotations on graphic designs~\cite{odonovan}, as well as mouse movements on natural images~\cite{jiang2015salicon}. 
We deployed \rev{10 experiments with 28 different parameter combinations on MTurk (Table~\ref{table-settings}): 7 experiments with information visualizations (testing 4 different bubble radius sizes on 3 different image subsets), 2 experiments with natural scenes (comparing mouse clicks to mouse movements), 7 experiments with static webpages (3 bubble radius sizes x 2 viewing times with free-viewing, and a separate description task), 1 experiment with graphic designs, and 11 experiments with another dataset of natural scenes (3 image blur sigmas x 3 bubble radius sizes with mouse clicks, and 2 blur sigmas with mouse movements)}.\\



\begin{table}[]
\centering
\tbl{Overview of BubbleView experiment settings including different image stimuli and parameters varied per experiment. We deployed a total of 10 experiments with 28 different parameter combinations. Bubble sigma and bubble radius are in pixels. Mouse modality corresponds to whether an image was revealed by discrete clicks, or by continuously moving the mouse cursor.}{
\begin{tabu} to 1.0\textwidth {X[1.40,c] X[0.15,c] X[0.45,c] X[0.5,c] X[0.45, c] X[0.3, c] X[0.4, c]}
\toprule[1.5pt]
\multirow{2}{*}{Dataset \& Image Type} & \multirow{2}{*}{Exp.} & \multicolumn{5}{c}{Experiment Parameters} \\ \cmidrule(l){3-7} 
 &  & Task type & Blur sigma & Bubble radius & Time (sec) & Mouse modality \\ \midrule
\multirow{3}{*}{\begin{tabular}[c]{@{}c@{}}MASSVIS~\cite{beyondmem}\\ Information visualizations\end{tabular}} & 1.1 & describe & 40 & 16, 24, 32 & unlim. & click \\ \cmidrule(l){2-7} 
 & 1.2 & describe & 40 & 24, 32, 40 & unlim. & click \\ \cmidrule(l){2-7} 
 & 1.3 & describe & 40 & 40 & unlim. & click \\ \midrule
\multirow{2}{*}{\begin{tabular}[c]{@{}c@{}}OSIE~\cite{xu2014predicting}\\ Natural scenes\end{tabular}} & 2.1 & free-view & 30 & 30 & 10 & click \\ \cmidrule(l){2-7} 
 & 2.2 & free-view & 30 & 30 & 5 & move \\ \midrule
\multirow{2}{*}{\begin{tabular}[c]{@{}c@{}}FIWI~\cite{shen2014webpage}\\ Static webpages\end{tabular}} & 3.1 & free-view & 50 & 30, 50, 70 & 10, 30 & click \\ \cmidrule(l){2-7} 
 & 3.2 & describe & 50 & 30 & unlim. & click \\ \midrule
\multirow{2}{*}{\begin{tabular}[c]{@{}c@{}}GDI~\cite{odonovan}\\ Graphic designs\end{tabular}} & \multirow{2}{*}{4} & \multirow{2}{*}{free-view} & \multirow{2}{*}{30} & \multirow{2}{*}{50} & \multirow{2}{*}{10} & \multirow{2}{*}{click} \\
 &  &  &  &  &  &  \\ \midrule
\multirow{2}{*}{\begin{tabular}[c]{@{}c@{}}SALICON~\cite{jiang2015salicon}\\ Natural scenes\end{tabular}} & 5.1 & free-view & 30, 50, 70 & 30, 50,  70 & 10 & click \\ \cmidrule(l){2-7} 
 & 5.2 & free-view & 30, 50 & 30 & 5 & move \\ \bottomrule
\end{tabu}}
\label{table-settings}
\end{table}

\subsection{Tasks and Procedures Overview}

We collected BubbleView data for 51 images selected out of each of 5 datasets (with additional images from the \nam{MASSVIS} dataset). 
We used two different tasks with the following instructions: 1) description: ``click and describe the image'', 2) free-viewing: ``click anywhere you want to look'' (Figure ~\ref{fig:interface}). The description task required at least 150 characters to ensure that participants completed the task with enough thoroughness. For the free-viewing task, the image description was not required but the time for viewing each image was \nam{fixed} to either 10 sec or 30 sec.
The description task is most appropriate for image types containing sufficient textual content to describe. 
The description task was used for information visualizations, while the free-viewing task was used for natural images, to make the BubbleView task instructions as close as possible to the original eye tracking experiments (\citeN{beyondmem} and \citeN{xu2014predicting}, respectively). We compared both task types on website images. 



The free-viewing task had 17 images per HIT, for a total of 3 different HITs to cover all 51 images (no overlap of images among HITs). With a 10 sec viewing time, a single HIT was timed to take 2.8 minutes to complete; with 30 sec of viewing time, 8.5 minutes.
 For the description task, since time per image was estimated to be significantly longer, there were only 3 images per HIT, for a total of 17 different HITs to cover all 51 images. No explicit time constraints were placed on this task. On average, the description HITs took about 9 minutes to complete.

To accept one of our HITs, a participant had to have an approval rate of over 95\% and live in the United States. After acceptance, the participant was asked to sign the informed consent before participating in the study. 
All participants were paid with approximately \$0.1/min rate which we translated to \$0.3 for the free-viewing task with 10 sec of viewing, \$0.9 for the free-viewing task with 30 sec of viewing, and \$0.5 for the description task\footnote{Our original estimates were that the description task would take 1.5 min per image, whereas in reality it took an average of 3.2 min per image. We later issued additional bonuses to compensate participants.}. All participants were paid regardless of whether they completed the task successfully or not. Some participant data was filtered out (see Supplemental Material) which is why the original number of participants recruited is not always equal to the final number of participants used for analysis.


\subsection{Analysis Overview} 
\label{sec:analysis_details}

Across all the experiments comparing BubbleView clicks to eye fixations we used the same set of analyses, which we describe here. We compared how well the distribution of BubbleView clicks approximates the distribution of eye fixations, using two metrics commonly used for saliency evaluation: Pearson's Correlation Coefficient (CC) and Normalized Scanpath Saliency (NSS) \cite{salMetrics_Bylinskii}. While the two metrics provide complementary evidence for our conclusions, the NSS metric also allows us to account for differences in \zoya{attentional consistency between participants (inter-observer congruency)} across datasets. \\

\noindent \textit{Converting clicks and fixations into maps} \\
Given a set of eye fixations on an image, we generate a \textbf{fixation map} by blurring the fixation locations with a Gaussian, with a sigma equal to one degree of the visual angle to approximate both the eye fovea and the measurement error of the eye tracker (a common evaluation choice~\cite{LeMeur2013,salMetrics_Bylinskii}). This produces a continuous map which, when properly normalized, can be interpreted as a 2D distribution containing the probability of participants looking at each image region. 
Similarly, given a set of BubbleView mouse clicks on an image, we compute a \textbf{BubbleView click map} by blurring the click locations with a Gaussian with the same sigma as for the ground truth fixation maps. \rev{We used a sigma of 10 for the OSIE dataset, and a sigma of 25 for the MASSVIS and FiWI datasets.} More generally, we refer to both fixation and click maps in this paper as \textbf{importance maps} for an image.
\\

\noindent \textit{Measuring the similarity between clicks and fixations} \\
We use two different metrics to measure the similarity scores between BubbleView clicks and eye fixations: Pearson's Correlation Coefficient (CC) and Normalized Scanpath Saliency (NSS). \rev{We compute similarity at the image level, by comparing the distributions of all clicks and fixations, across participants, per image. We then average all the per-image scores to obtain the similarity scores for a dataset.}

To obtain the \textbf{CC score} for an image, we measure how well the click map predicts the fixation map, as a correlation between the two maps (see the Supplemental Material for details). The CC score is 0 when the two maps are not correlated, and 1 when they are identical. To obtain the \textbf{NSS score} for an image, we measure how well the click map predicts the discrete fixation locations (in this case, we do not compute a fixation map). We compute the average click map value at the fixated locations, after normalizing the click map. A map that is at chance at predicting fixation locations would receive an NSS score of 0, while a positive NSS indicates predictive power. 

The advantage of the CC score is that it is bounded between 0 and 1, and can provide a simple, interpretable summary score that is ambivalent to the number of fixations that were used to generate the fixation map. The advantage of the NSS score is that it is computable for different numbers of eye tracking participants, and we use it for finer-grained analyses to examine how performance changes as we increase the number of participants. NSS is not bounded; to turn NSS into a bounded score, we can normalize it by inter-observer consistency, as described below.\\


\noindent \textit{Accounting for inter-observer consistency} \\
If different eye tracking participants look at different regions of the image, they can not be used to predict each other's fixations. In these cases, BubbleView clicks will also not be as predictive of the fixations. For a fair evaluation, we normalize the BubbleView scores by the consistency of the eye tracking participants in a given dataset.

Consistency between eye tracking participants is measured in the following way:
the fixations of all but one observer (i.e., N-1 observers) are aggregated into a fixation map which is used to predict the fixations of the remaining observer. This is repeated by leaving out one observer at a time, and then averaging the prediction performance to obtain the resulting inter-observer congruency (\textbf{IOC}) or inter-subject consistency \cite{Borji_2012,wilming2011measures,LeMeur2013}. We measure IOC using the NSS metric. 


We first compute the NSS score of the BubbleView click map at predicting all the eye fixations collected on an image, across all the observers. Then we normalize this score by the IOC of the eye tracking participants on that dataset. The resulting \textbf{normalized NSS} score can be interpreted as: the percent of the eye fixations accounted for, or predicted by, the BubbleView clicks. \\

\noindent \textit{Measuring performance in the limit} \\
We consider performance when the number of study participants is taken to the limit, to get an upper bound on performance and determine if any systematic differences exist between methodologies that can not be reduced by gathering more data. To do this, we measure the ability of BubbleView click maps to predict ground-truth fixation locations, for different numbers of BubbleView participants. We obtain an NSS score for different numbers of participants $n$\rev{, by randomly selecting $n$ participants for each of 10 splits, and averaging the results.}
Then we fit these scores to the power function $f(n) = a*n^b + c$, constraining $b$ to be negative. Taking $n$ to the limit, $c$ is the NSS score at the limit. 
In cases where the total number of BubbleView participants for a particular experiment is not enough for a robust model fitting, we omit this analysis. 

\section{Experiments comparing BubbleView clicks to eye fixations}
\subsection{Experiment 1: comparison to eye fixations on information visualizations}\label{sec:exp1}

We began by exploring how well BubbleView clicks on information visualizations gathered on MTurk approximate eye fixations collected in a controlled lab setting. In the initial experiments in \citeN{kim2015crowdsourced}, we had gathered BubbleView data on 51 visualizations with a bubble radius size of 16 pixels. Here we extended these experiments to explore the effect of bubble radius size and number of participants on the quality of BubbleView data. We varied the bubble radius between 16 and 40 pixels, and collected up to 40 participants worth of clicks per image.\\ 

\begin{myframe}

\textit{Motivating questions}\vspace{-0.5em}
\begin{itemize}[label=\ding{212}]
\item How does bubble radius size affect performance?
\item How many BubbleView participants is enough?
\end{itemize}
\end{myframe}

\noindent \textit{Stimuli} \\
The MASSVIS dataset contains over 5,000 information visualization images, of which 393 ``target'' images contain the eye movements of 33 participants free-viewing each image for 10 seconds as part of a memory test at the end of the study \cite{beyondmem}. In the eye tracking set-up, images were shown full-screen with a maximum dimension of 1000 pixels to a side, where 1 degree of viewing angle corresponded to 32.6 pixels. \zoya{Participants made on average 39 fixations per image, or 3.9 fixations/sec.}

\begin{figure}
 \centering
\includegraphics[width=1\textwidth]{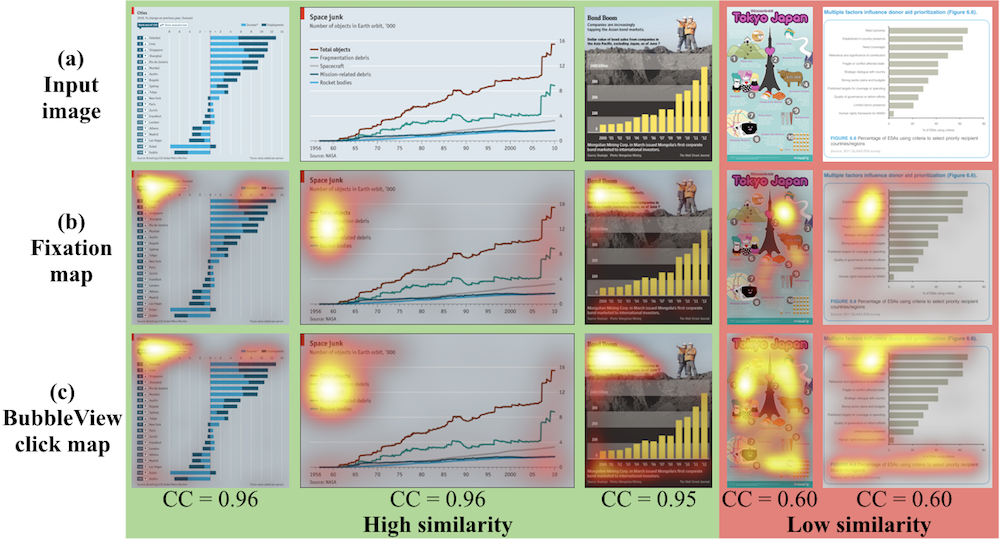}
 \caption{Example images from the MASSVIS dataset. Dataset images (a), with corresponding ground-truth fixation maps (b) and BubbleView click maps (c). We show cases where BubbleView maps have high similarity, and cases with low similarity, to fixation maps.}
 \label{fig:datasets1}
\end{figure}

We selected 202 from the total 393 target images, spanning infographic, news media, and government publication categories (Figure~\ref{fig:datasets1}). We chose visualizations that had sufficiently large text and enough context to understand them without requiring specialized knowledge. We resized the images to half their original size with a maximum dimension of 500 pixels to a side. 
The images were blurred with a sigma of 40 pixels, which we found distorted the text in these images beyond legibility \cite{beyondmem,kim2015crowdsourced}. \\

\noindent \textit{Method} \\
We ran a series of experiments to progressively find a bubble radius that best approximates eye fixations: \textbf{Exp. 1.1} with one set of 51 images and bubble radius sizes of 16, 24, and 32 pixels respectively, \textbf{Exp. 1.2} with another set of 51 images and bubble radius sizes of 24, 32, and 40 pixels, and \textbf{Exp. 1.3} with the remaining 100 images with a bubble radius of 32 pixels, which we determined from the first two experiments to produce good data quality. A bubble radius of 32 pixels corresponds to about 2 degrees of visual angle in the eye tracking studies on the original-sized images. 

In a single HIT, participants were shown a random sequence of 3 images, and asked to describe each image with no time constraints on the task, allowing for individual differences in the time to write image descriptions. 

For Exp. 1.1, we requested enough HITs so that each image would be seen by an average of 40 participants. From this experiment we found that 10--15 participants are sufficient for \rev{achieving high similarity scores to eye fixations}, and proceeded to collect an average of 10--15 participants for each image in Exp. 1.2 and Exp. 1.3.\\


\noindent \textit{Results on bubble size} 

\zoya{Participants explored each image for an average of 3 minutes, iterating between clicking around and typing text. As bubble size increased, the number of clicks and total task time monotonically decreased. Participants made an average of 103 clicks per image (0.5 clicks/sec) with a bubble radius of 16 pixels, 65 clicks (0.3--0.4 clicks/sec) with a bubble radius of 32, and 55 clicks (0.3 clicks/sec) with a bubble radius of 40 pixels. Depending on the bubble size, participants spent 15--30\% of the task time clicking, and the rest of the time typing a description.} After receiving a number of participant complaints about task difficulty at a bubble radius of 16 pixels, we discontinued the use of this bubble radius in future experiments.

\begin{figure}[h]
 \centering
\includegraphics[width=1\textwidth]{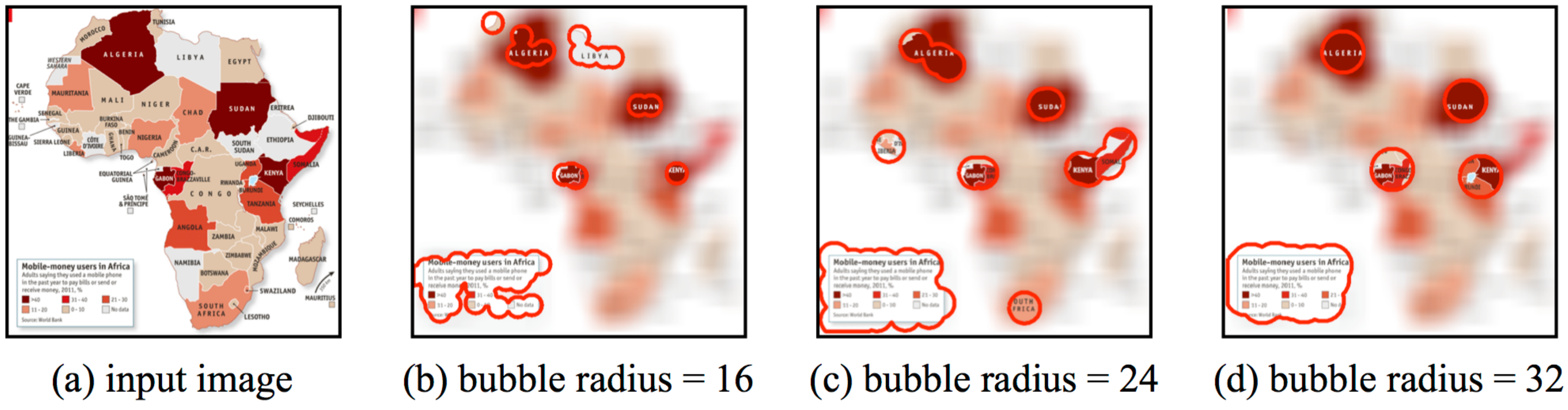}
 \caption{We found few differences in the resulting click maps from different settings of the bubble radius. Plotted here are the clicks of 3 participants (b-d) who explored the same image (a) with BubbleView, but with a different bubble size: 16, 24, and 32 pixel radius, respectively. The smaller the bubble, the more clicks a participant made, and the longer the task took to complete. Overall, the same regions of interest tended to be clicked on, despite differences in bubble sizes.}
 \label{fig:difbubblesizes}
\end{figure}

We computed the similarity between the BubbleView click maps and ground truth fixation maps across all images for all settings of bubble radius (Table~\ref{tab:massvis}). To make scores comparable, we set the number of participants $n=10$ when computing the BubbleView click maps (the common denominator across all experiments). 
The similarities between the BubbleView click maps and the fixation maps were close across all bubble radius sizes (CC = 0.82--0.86). \rev{Because the different subsets of the MASSVIS dataset used in Exp. 1.1--1.3 had different inter-observer consistency (IOC) values\footnote{This is an artifact of the images being different in the different subsets. In particular, Exp. 1.1 ended up containing more news media images and less government and infographic images than Exp. 1.2.}, normalized NSS scores are more comparable across experiments than raw NSS scores.}
The normalized NSS score was very similar across all bubble radius sizes, with BubbleView clicks accounting for an average of 89--90\% of eye fixations with 10 participants, and \rev{climbing up to $92\%$ for larger numbers of participants ($n \geq 18$).}
\rev{Running a one-way ANOVA with bubble size as the factor, we did not find any significant effects of radius size on the similarity of clicks to fixations, under either of CC and NSS scores ($F<1$ for all comparisons). Although the number of clicks changed, the overall pattern of bubble clicks remained the same (Figure~\ref{fig:difbubblesizes}).}


\begin{myframe}
\textit{Take-aways:} no significant differences were found between bubble sizes in terms of similarity of BubbleView clicks to eye fixations. Bubble sizes in the range 24--40 pixels were found appropriate. Smaller bubble sizes increased the task time and effort.
\end{myframe}

\begin{figure}
 \centering
\includegraphics[width=0.6\textwidth]{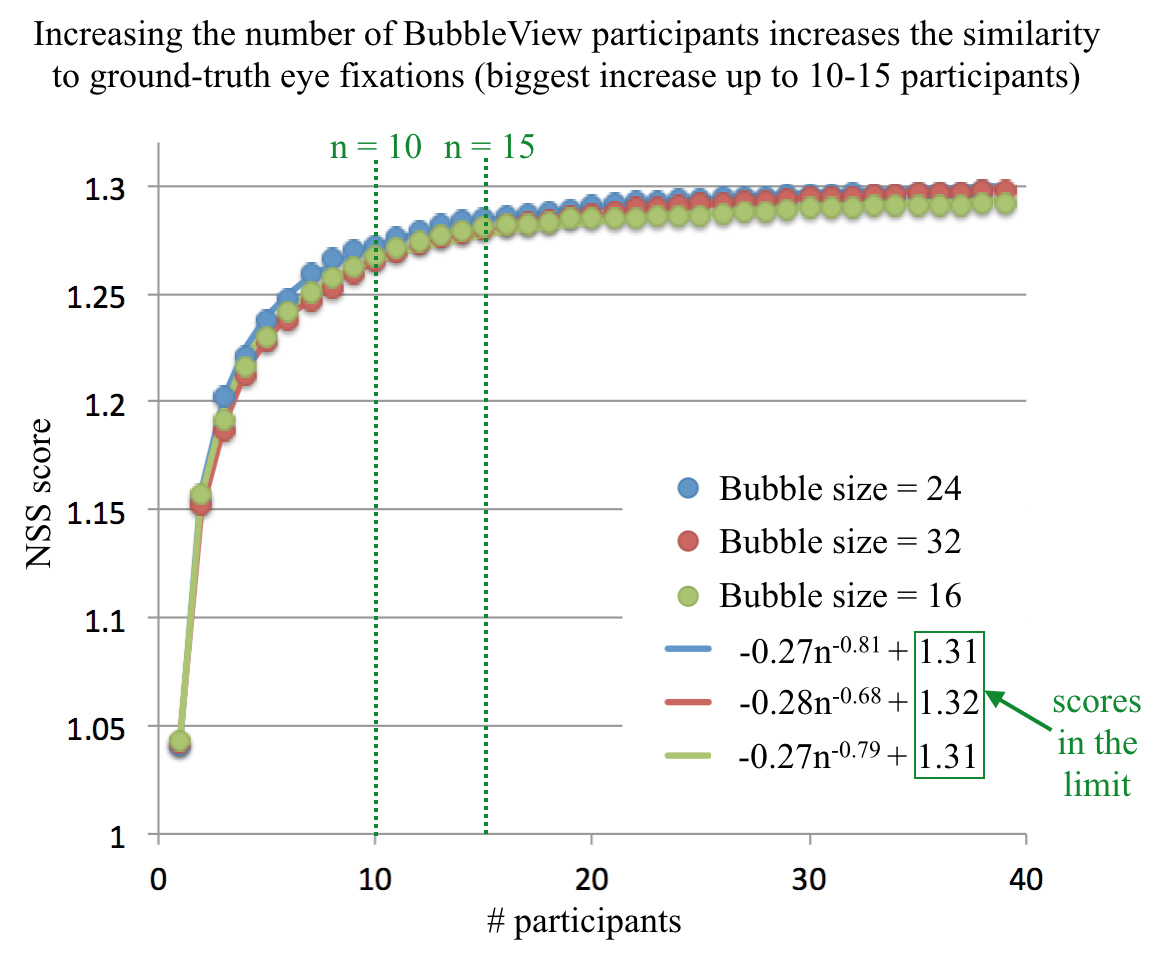}
 \caption{The NSS score of BubbleView click maps computed with different numbers of participants, when used to predict discrete fixation locations on the MASSVIS dataset. Each point represents the score obtained at a given number of participants\rev{, averaged over 10 random splits of participants,} and all 51 images used in Exp. 1.1. We include data points from 3 different bubble radius sizes. By fitting power functions of the form $an^b + c$ to each set of points, we find that these scores do not change significantly in the limit of participants ($n \rightarrow \infty$).}
 \label{fig:bubble_dif_participants}
\end{figure}

\noindent \textit{Results on number of participants} \\
In Exp. 1.1 we collected an average of 40 participants of BubbleView clicks per image to investigate how BubbleView maps change with the number of participants (Figure~\ref{fig:bubble_dif_participants}).
\rev{As described in Section~\ref{sec:analysis_details}, we fit power functions to the NSS scores for different numbers of participants to extrapolate performance.}
We found that after about 10--15 participants, the similarity of BubbleView click maps to ground truth fixation maps was already 97-98\% of the performance achievable in the limit. The NSS score was extrapolated to increase to 1.31 in the limit \rev{(95\% C.I. [1.312, 1.315])}
with a bubble size of 16, 1.32 in the limit \rev{(95\% C.I. [1.320, 1.324])} with a bubble size of 24, and 1.31 in the limit \rev{(95\% C.I. [1.306, 1.310])} with a bubble size of 32. As a result of these analyses, we used an average of 10--15 participants for all future BubbleView experiments. 

\begin{myframe}
\textit{Take-aways:} 10--15 participants worth of BubbleView clicks already accounted for up to 97-98\% of the performance achievable in the limit of the number of participants.
\end{myframe}

\ignore{ MASSVIS
}
\begin{table}[]
\centering

\tbl{ We evaluated BubbleView clicks at approximating ground-truth eye fixations on the MASSVIS dataset by varying the bubble radius. We ran 3 sets of experiments on different subsets of the MASSVIS dataset. We measured the cross-correlation (CC) between BubbleView click maps and ground truth fixation maps, averaged over all images (CC has an upper bound of 1). The normalized scanpath saliency (NSS) score measured how well BubbleView click maps predict discrete fixation locations, averaged over all images. The NSS upper bound depends on the ground-truth data, so we included the inter-observer consistency (IOC) score of the eye tracking participants (measured using NSS). Normalizing the NSS score of the BubbleView maps by IOC allows us to report the percent of ground-truth fixations predicted by the BubbleView maps. To make the scores comparable across all the experiments, we fixed the number of participants to $n=10$. \rev{In gray we report the results obtained by including all $n$ participants that were collected for each experiment. 
The difference in CC and NSS scores with different bubble radius sizes was not significant ($F<1$ for all comparisons).}
}
{
\begin{tabu} to 1.0\textwidth{X[1.5,c] X[1.0,c] X[0.5,c] X[0.5,c] X[1.0,c]}
\toprule
\multicolumn{1}{c}{Exp. 1: visualizations} & \multicolumn{1}{c}{Bubble Radius (pixel)} & \multicolumn{1}{c}{CC} & \multicolumn{1}{c}{NSS} & \multicolumn{1}{c}{Normalized NSS} \\ \midrule\midrule
\multirow{3}{*}{\begin{tabular}[c]{@{}l@{}}Exp. 1.1: 51 visualizations \\ Description task\\ (ground-truth IOC: 1.42)\end{tabular}} & 16                    & \begin{tabular}[c]{@{}l@{}}0.86 \\ \color{gray}{0.87}\end{tabular} & \begin{tabular}[c]{@{}l@{}}1.27 \\ 
\color{gray}{1.30} \end{tabular} & \begin{tabular}[c]{@{}l@{}}89\% ($n=10$) \\ \color{gray}{92\% ($n=38$)}\end{tabular} \\ \cline{2-5}
                                                                                                                                      & 24                    & \begin{tabular}[c]{@{}l@{}}0.86 \\ \color{gray}{0.87}\end{tabular} & \begin{tabular}[c]{@{}l@{}}1.27\\ \color{gray}{1.30}\end{tabular}  & \begin{tabular}[c]{@{}l@{}}89\% ($n=10$) \\ \color{gray}{92\% ($n=39$)} \end{tabular} \\ \cline{2-5}
                                                                                                                                      & 32                    & \begin{tabular}[c]{@{}l@{}}0.86 \\ \color{gray}{0.87}\end{tabular} & \begin{tabular}[c]{@{}l@{}}1.27\\ \color{gray}{1.29} \end{tabular}  & \begin{tabular}[c]{@{}l@{}}89\% ($n=10$) \\ \color{gray}{91\% ($n=40$)} \end{tabular} \\\midrule
\multirow{3}{*}{\begin{tabular}[c]{@{}l@{}}Exp. 1.2: 51 visualizations \\ Description task\\ (ground-truth IOC: 1.33)\end{tabular}} & 24                    & \begin{tabular}[c]{@{}l@{}}0.82 \\ \color{gray}{0.84}\end{tabular} & \begin{tabular}[c]{@{}l@{}}1.20\\ \color{gray}{1.22} \end{tabular}  & \begin{tabular}[c]{@{}l@{}}90\% ($n=10$)\\ \color{gray}{92\% ($n=20$)} \end{tabular} \\ \cline{2-5}
                                                                                                                                      & 32                    & \begin{tabular}[c]{@{}l@{}}0.84 \\ \color{gray}{0.85}\end{tabular} & \begin{tabular}[c]{@{}l@{}}1.20\\ \color{gray}{1.22}\end{tabular}  & \begin{tabular}[c]{@{}l@{}}90\% ($n=10$) \\ \color{gray}{92\% ($n=18$)}\end{tabular} \\ \cline{2-5}
                                                                                                                                      & 40                    & \begin{tabular}[c]{@{}l@{}}0.83 \\ \color{gray}{0.84}\end{tabular} & \begin{tabular}[c]{@{}l@{}}1.19 \\ \color{gray}{1.19} \end{tabular}  & \begin{tabular}[c]{@{}l@{}}89\% ($n=10$) \\ \color{gray}{89\% ($n=11$)} \end{tabular} \\\midrule
\begin{tabular}[c]{@{}l@{}}Exp. 1.3: 100 visualizations \\ Description task \\ (ground-truth IOC: 1.35)\end{tabular}              & 32                    & \begin{tabular}[c]{@{}l@{}}0.84 \\ \color{gray}{0.84}\end{tabular} & \begin{tabular}[c]{@{}l@{}}1.21\\ \color{gray}{1.21} \end{tabular}  & \begin{tabular}[c]{@{}l@{}}90\% ($n=10$)\\ \color{gray}{90\% ($n=10$)}\end{tabular} \\ \bottomrule
\end{tabu}
}
\label{tab:massvis}
\end{table}

\begin{figure}
 \centering
a)\includegraphics[height=6.5cm]{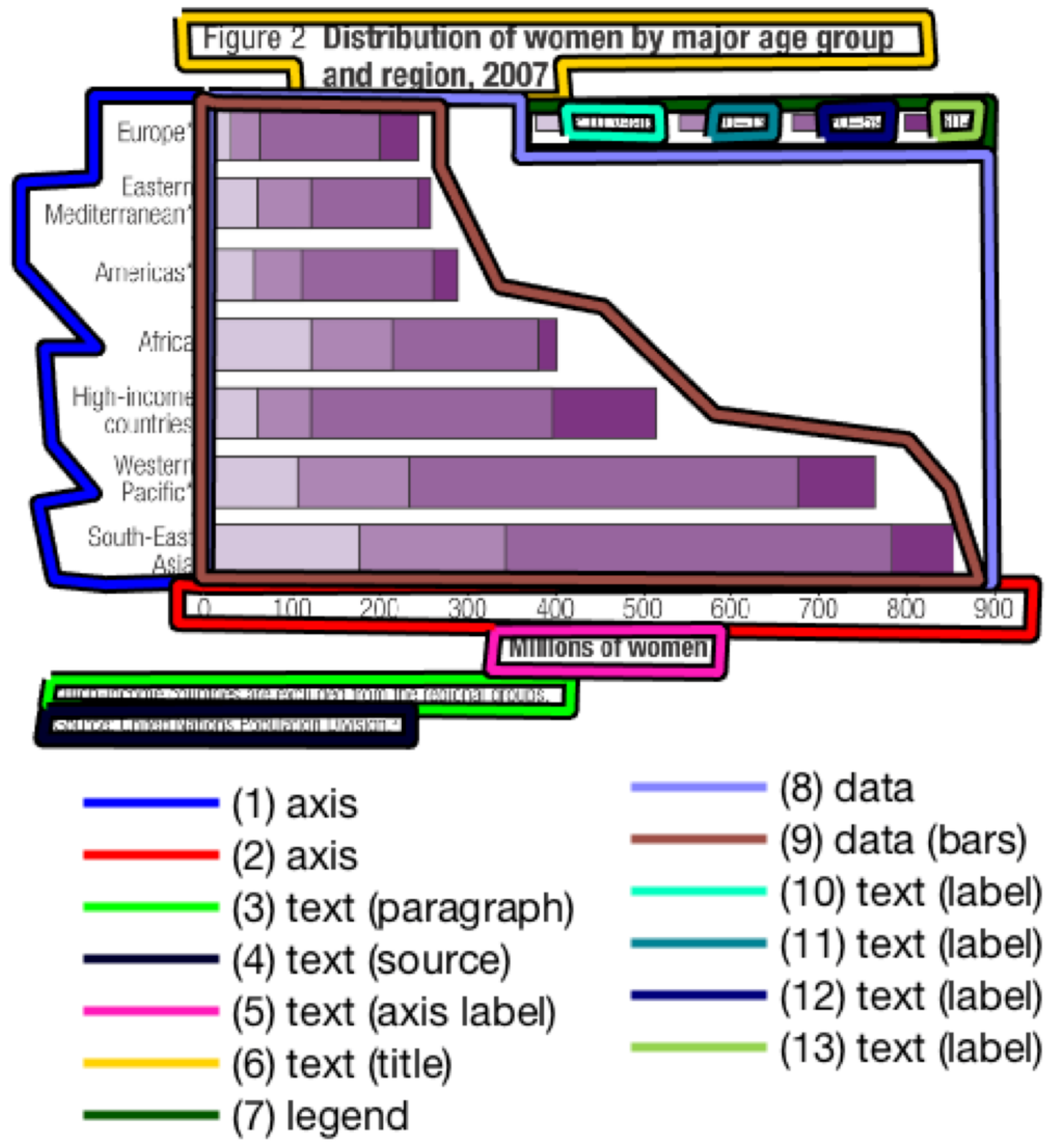}
b)\includegraphics[height=6.5cm]{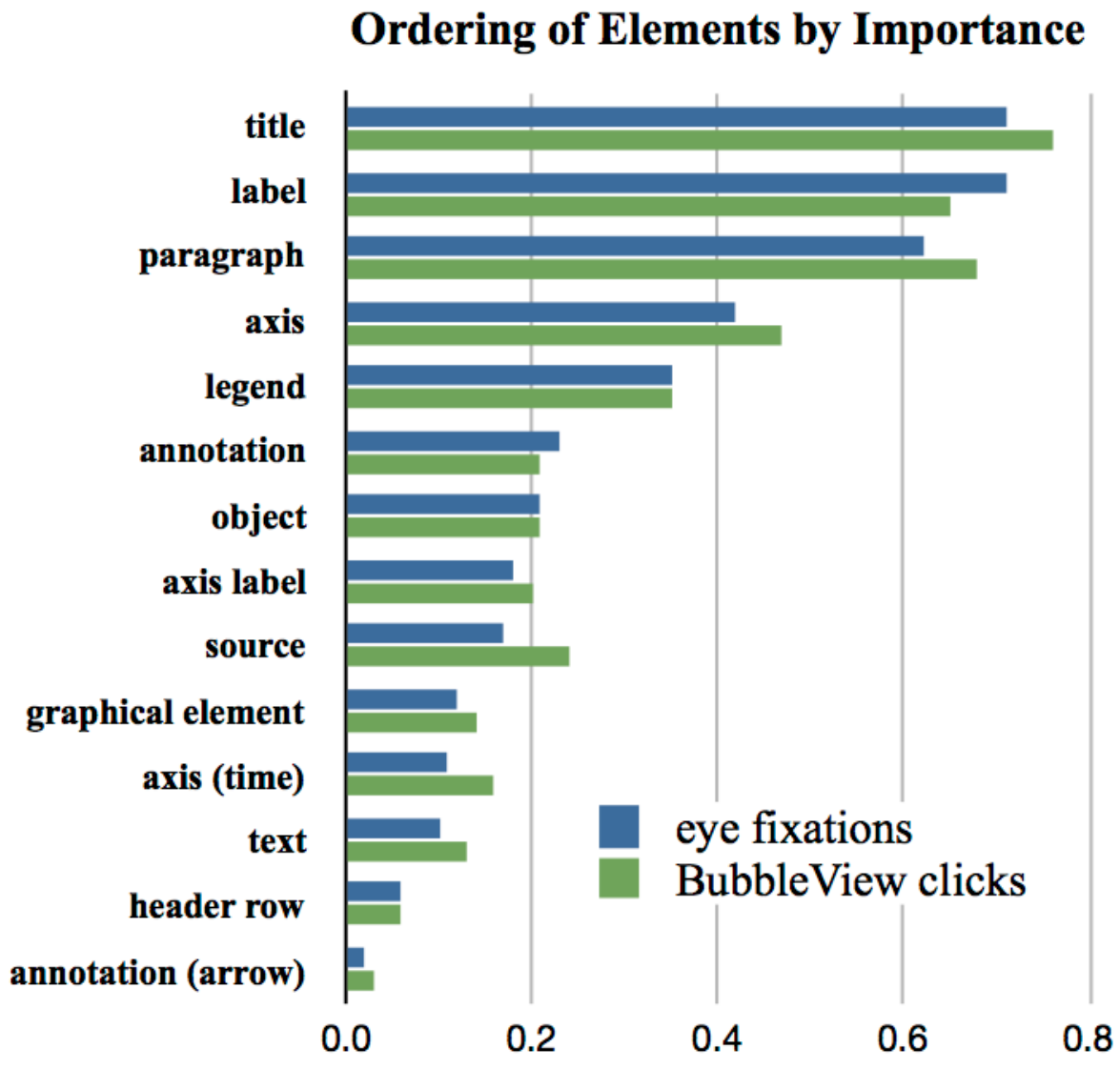}
 \caption{(a) An example of a labeled visualization from the MASSVIS dataset. (b) By overlapping fixation maps and BubbleView click maps with such element annotations (and taking the maximum value of the map inside the element), we obtain an importance score for each element in each visualization. By averaging across 202 visualizations, we obtain an aggregate importance score per element type.
}
 \label{fig:viselements}
\end{figure}

\noindent \textit{Results on ranking elements by importance} \\
We also explored the relationship between BubbleView clicks and eye fixations at ranking visualization elements by importance. For this purpose we used the element segmentations (e.g., title, axis, legend, etc.) available in the MASSVIS dataset~\cite{beyondmem}. For each of the 202 visualizations from Exp. 1, we overlapped the element segmentations with the fixation map of the visualization, and took the maximum value of the fixation map within the element's boundaries as its \textbf{importance score}, as in~\cite{jiang2015salicon,bylinskii2016should}. We averaged the element scores across all 202 visualizations to obtain an aggregate importance score for each type of element (Figure~\ref{fig:viselements}). 
We repeated this computation using the BubbleView click maps of the visualizations to get another set of importance scores for the same elements. The ranking of elements by importance scores according to BubbleView clicks is highly correlated to the ranking according to eye fixations (Spearman correlation = 0.96).  

\begin{myframe}
\textit{Take-aways:} BubbleView can be used to rank visualization elements by importance, predicting how often people would fixate those elements during natural viewing.
\end{myframe}

\subsection{Experiment 2: comparison to eye fixations on natural images}\label{sec:exp2}

In Experiment 1 we found that BubbleView clicks offered a very good approximation to eye fixations on information visualizations with a description task.
However, because free-viewing is a more common setting for human perception studies \rev{of natural images (specifically for saliency datasets)}, we wanted to determine if BubbleView clicks can also be used to approximate free-viewing fixations on natural images. We used similar BubbleView settings to the ones found in Exp. 1: a bubble size of 30 pixels and 15 participants worth of clicks.

\begin{myframe}

\noindent \textit{Motivating questions}\vspace{-0.5em}
\begin{itemize}[label=\ding{212}]
\item Does BubbleView generalize to natural images with a free-viewing task?
\end{itemize}

\end{myframe}

\begin{figure}[h]
 \centering
\includegraphics[width=1\textwidth]{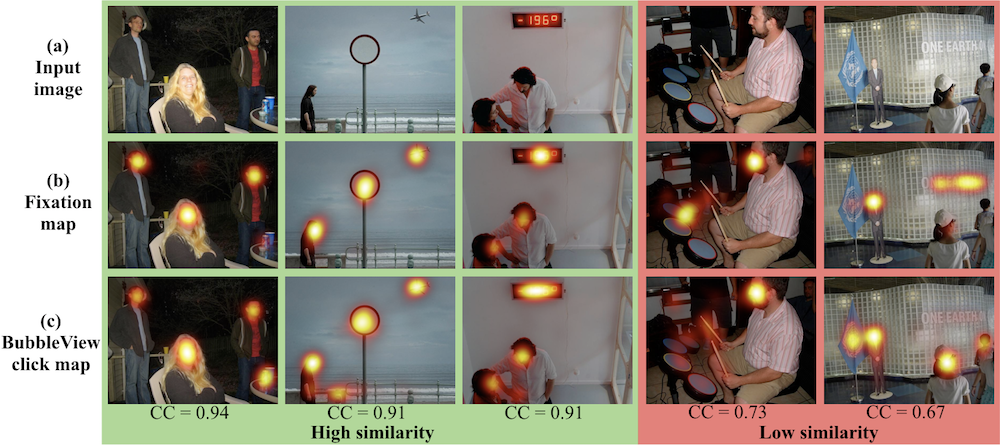}
 \caption{Example images from the OSIE dataset. Dataset images (a), with corresponding ground-truth fixation maps (b) and BubbleView click maps (c). We show cases where BubbleView maps have high similarity, and cases with low similarity, to fixation maps.}
 \label{fig:datasets2}
\end{figure}

\noindent \textit{Stimuli} \\
The OSIE dataset contains 700 natural images with multiple dominant objects per image \cite{xu2014predicting}.  Eye movements on this dataset were collected by instructing 15 participants to free-view each image for 3 seconds. \zoya{Participants made an average of 9.3 fixations per image (3.1 fixations/sec).} In this eye tracking setup, images were presented at a resolution of $800\times600$ pixels and 1 degree of viewing angle corresponded to 24 pixels. 
For our study, we randomly sampled 51 OSIE images (Figure~\ref{fig:datasets2}), downsized them to $640\times480$ pixels, and blurred them with a sigma of 30 pixels. \\ 

\noindent \textit{Method} \\
In \textbf{Exp. 2.1}, we asked participants to free-view a series of images and to click anywhere they want to look for 10 sec per image. We used a bubble radius of 30 pixels, equal to about 1.5 degrees of visual angle in the eye tracking study. Although the viewing time for the OSIE eye tracking study was 3 sec per image, we increased this time for the BubbleView experiment to account for the time of clicking a mouse. We piloted different viewing times and determined 10 sec to be appropriate \rev{(clicking took about 3 times as long as natural viewing)}. We collected an average of \rev{60} participants worth of BubbleView click data for each image.

\rev{Apart from ground truth eye fixations, mouse movements using the related SALICON methodology are also available for the OSIE dataset~\cite{jiang2015salicon}. To facilitate a direct comparison between BubbleView and SALICON, in \textbf{Exp. 2.2} we re-ran data collection with BubbleView, replacing mouse clicks with mouse movements, with a bubble radius of 30 pixels. As in SALICON, we used a task time of 5 seconds.
The results of this experiment are discussed in Section~\ref{sec:exp5}, in the context of other comparisons to the SALICON methodology.} \\


\noindent \textit{Results} \\
\zoya{During 10 seconds of viewing, participants made an average of 13.1 clicks, or about 1.3 clicks/sec - three times fewer clicks than fixations per second.
} 

In Exp. 2.1, the similarity between BubbleView click maps and ground truth fixation maps with free-viewing on natural images was smaller (NSS = 2.61, CC = 0.81, Table~\ref{tab:osie}) than in Exp. 1 with visualizations. Even though eye tracking participants are quite consistent with each other on the OSIE dataset (IOC = 3.35), BubbleView participants are not as predictive of eye tracking participants in this case. 
\rev{BubbleView clicks of 54 participants can predict 80\% of eye fixations, while the projected performance in the limit only converges to 82\% (95\% C.I. [2.742, 2.754])}. However, 10 BubbleView participants can already account for 78\% of eye fixations. 

Exp. 2.2 showed that a related methodology using a moving-window approach \cite{jiang2015salicon} is no better at approximating ground-truth eye fixations on this dataset (Table~\ref{tab:osie}). In fact, to achieve the same performance as BubbleView, SALICON actually requires more participants (Section~\ref{sec:exp5}).
BubbleView can serve as \nam{an affordable and scalable} alternative. When running a large number of eye tracking experiments is infeasible, BubbleView can be used for studying human perception and collecting large-scale saliency datasets (as in \citeN{jiang2015salicon}; see \citeN{predimportance}). 

\begin{myframe}
\textit{Take-aways:} Similarity between BubbleView clicks and eye fixations is lower on natural images with a free-viewing task than with visualizations with a description task. Despite this, 10 BubbleView participants can already account for 78\% of eye fixations on natural images, so BubbleView can still serve as an affordable approximation to eye tracking.
\end{myframe}

\begin{table}[h]
\centering
\tbl{We evaluated BubbleView clicks at approximating ground-truth eye fixations on the OSIE dataset. We ran BubbleView data collection using mouse clicks (Exp. 2.1) and using mouse movements (Exp. 2.2). For comparison, we also include the performance of the SALICON methodology on the same dataset (Section~\ref{sec:exp5}).
For fair comparison with in-lab SALICON, we only used $n=12$ participants per image per study. \rev{The difference in scores at $n=12$ participants was not significant [F(200)=1.81, n.s.]. In gray we report the results obtained by including all $n$ participants that were collected for each experiment.} 
}
{
\begin{tabu}{X[2.0, c] X[0.5, c] X[0.5, c] X[1.0, c]}
\toprule
\multicolumn{1}{c}{Exp. 5.3: natural scenes (ground-truth IOC: 3.35)} & \multicolumn{1}{c}{CC} & \multicolumn{1}{c}{NSS} & \multicolumn{1}{c}{Normalized NSS} \\ \midrule\midrule
BubbleView (clicks) & \begin{tabular}[c]{@{}l@{}}0.81\\ \color{gray}{0.84} \end{tabular} & \begin{tabular}[c]{@{}l@{}}2.61\\ \color{gray}{2.69} \end{tabular} & \begin{tabular}[c]{@{}l@{}}78\% ($n=12$)\\ \color{gray}{80\% ($n=54$)} \end{tabular} \\ \midrule
BubbleView (movements) & \begin{tabular}[c]{@{}l@{}}0.81\\ \color{gray}{0.83} \end{tabular} & \begin{tabular}[c]{@{}l@{}}2.52\\ \color{gray}{2.55} \end{tabular} & \begin{tabular}[c]{@{}l@{}}75\% ($n=12$)\\ \color{gray}{76\% ($n=49$)} \end{tabular} \\ \midrule
SALICON & \begin{tabular}[c]{@{}l@{}}0.81\\ \color{gray}{0.84} \end{tabular} & \begin{tabular}[c]{@{}l@{}}2.52\\ \color{gray}{2.61} \end{tabular} & \begin{tabular}[c]{@{}l@{}}75\% ($n=12$)\\ \color{gray}{78\% ($n=92$)} \end{tabular} \\ \midrule
In-lab SALICON & \begin{tabular}[c]{@{}l@{}}0.81\\ \color{gray}{0.81} \end{tabular} & \begin{tabular}[c]{@{}l@{}}2.61\\ \color{gray}{2.61} \end{tabular} & \begin{tabular}[c]{@{}l@{}}78\% ($n=12$)\\ \color{gray}{78\% ($n=12$)} \end{tabular} \\ \bottomrule
\end{tabu}
}
\label{tab:osie}
\end{table}

\subsection{Experiment 3: comparison to eye fixations on static webpages}\label{sec:exp3}

Apart from natural images, webpages are another image type that frequently serve as the focus of eye tracking and usability studies \cite{shen2014webpage,shen2015predicting,buscher2009you,nielsen2010eyetracking,rodden2008eye,chen2001can}.
For this reason, we wanted to test the generalizability of the BubbleView methodology to webpages. Because the static webpage images were denser in visual and information content than the information visualizations and natural images from the first two experiments, we evaluated a number of different BubbleView settings to try to find the best approximation to eye fixations. We varied bubble radius size and viewing time. \rev{As in the original FiWI eye-tracking experiment, we started with a free-viewing task.} Similar to Exp. 1, we also tried a description task with unlimited task time. 


\begin{myframe}
\noindent \textit{Motivating questions}\vspace{-0.5em}
\begin{itemize}[label=\ding{212}]
\item Does BubbleView generalize to webpages?
\item How do the task and viewing time affect performance?
\item Does viewing time interact with bubble size?
\end{itemize}
\end{myframe}

\noindent \textit{Stimuli} \\
The FiWI dataset contains 149 screenshots of static webpages collected from various sources on the Internet and sorted into pictorial (dominated by pictures such as photo sharing websites), text (high density text such as encyclopedia websites), and mixed types \cite{shen2014webpage}. Eye movements on this dataset were collected by instructing 11 participants to free-view each webpage for 5 seconds. \zoya{Participants made an average of 17.9 fixations per image (3.6 fixations/sec).} In this eye tracking setup, 1 degree of visual angle was approximately 50 pixels. 

\begin{figure}
 \centering
\includegraphics[width=1\textwidth]{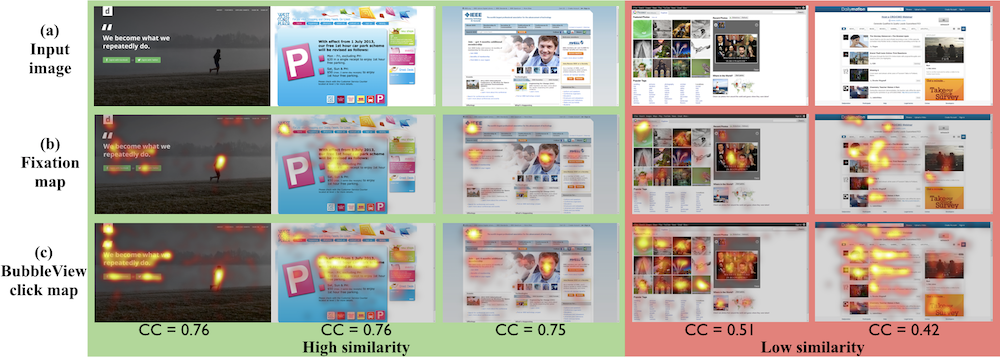}
 \caption{Example images from the FiWI dataset. Dataset images (a), with corresponding ground-truth fixation maps (b) and BubbleView click maps (c). We show cases where BubbleView maps have high similarity, and cases with low similarity, to fixation maps.}
 \label{fig:datasets3}
\end{figure}

We sampled 17 images from each of the three categories (pictorial, text, mixed), resulting in a total of 51 images (Figure~\ref{fig:datasets3}). We downsized the images from $1360\times768$ pixels to $1000\times565$ pixels to fit within a typical MTurk browser window, while preserving image aspect ratios. These webpages tended to have more varied font size compared to the images in Exp. 1--2. We manually selected a blur sigma of 50 pixels to distort the text on these images beyond legibility. \\ 

\noindent \textit{Method} \\
We ran experiments with two task types where participants were asked to either free-view or describe each webpage. In \textbf{Exp. 3.1}, with the free-viewing task, we used a 2 x 3 factorial design (viewing time: 10 sec or 30 sec; bubble radius: 30, 50, or 70 pixels). In \textbf{Exp. 3.2}, with the description task, we used a bubble radius of 30 pixels and unlimited time. We collected an average of 15 participants worth of BubbleView click data for each image under each task.\\

\begin{figure}
 \centering
\includegraphics[width=0.7\textwidth]{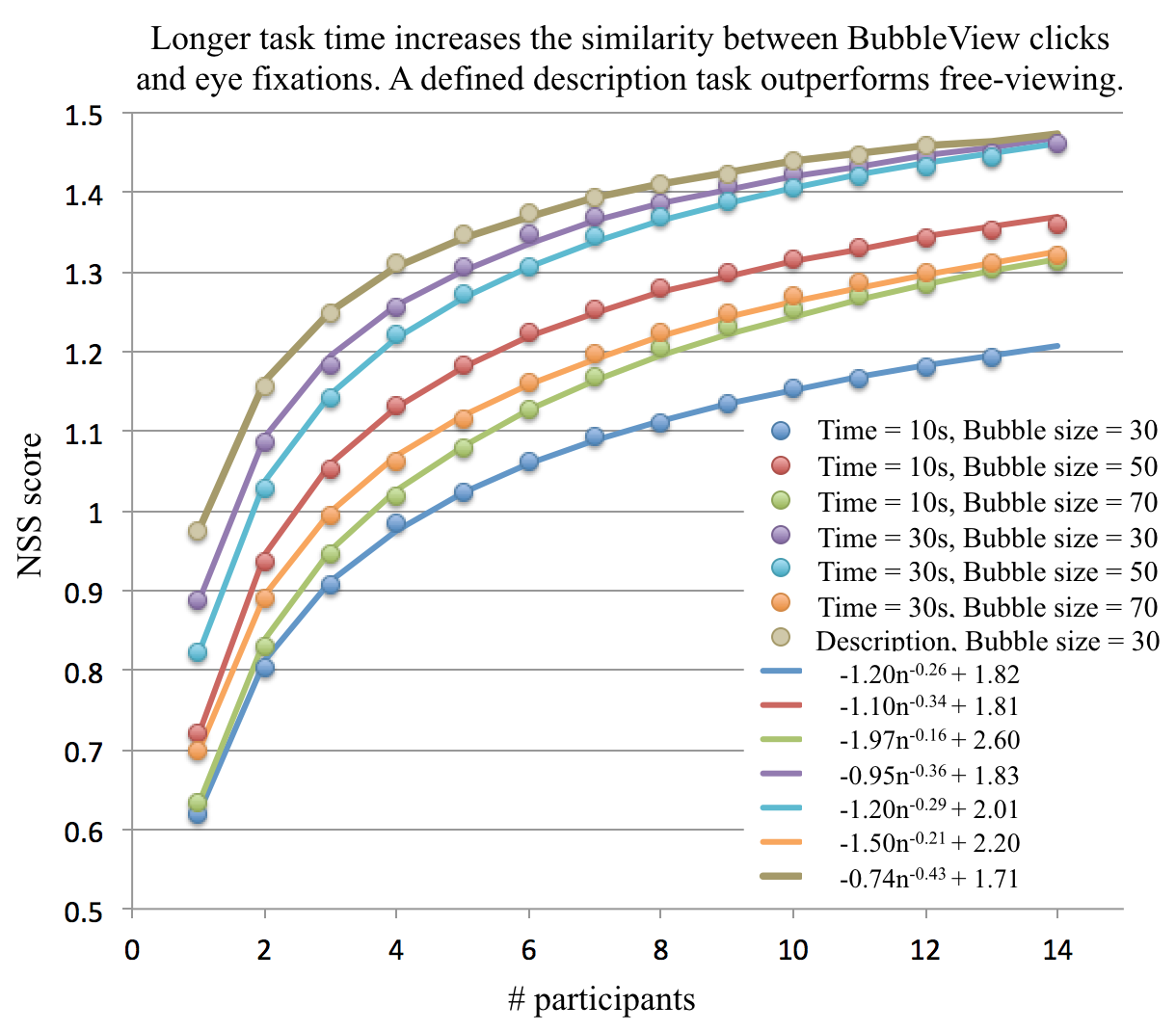}
 \caption{The NSS score of BubbleView click maps computed with different numbers of participants, when used to predict discrete fixation locations on the FiWI dataset. Each point represents the score obtained at a given number of participants, averaged over \rev{10 random splits of participants, and} all 51 images.
 }
 \label{fig:websaly_graphs}
\end{figure}

\noindent \textit{Results on stimuli} \\
\zoya{In the free-viewing task (Exp. 3.1), participants made an average of 1.0--1.8 clicks/sec, while in the description task (Exp. 3.2), participants made an average of 0.5 clicks/sec, indicating that they spent more than half the time typing descriptions. Clicking took about 3 times longer than natural viewing. As in Exp. 1, the number of clicks per second monotonically decreased with increasing bubble size, even though viewing time was fixed (Exp. 3.1). Tripling the viewing time from 10 to 30 seconds did not quite triple the number of clicks, but increased them by 2.2--2.6 times.}

The similarity between BubbleView click maps and ground truth fixation maps on webpages was lowest of all image types tested so far in Exp. 1--3 (Table~\ref{tab:fiwi}). However, the inter-observer consistency of eye tracking participants is also lowest on webpages (IOC = 1.85). Recall that IOC between eye tracking participants serves as an upper bound for how well BubbleView clicks can predict eye fixations. After accounting for IOC, the normalized NSS scores show that BubbleView clicks can account for up to 78\% of eye fixations on webpages, similar to the score on natural images (Exp.~2).

IOC was highest on the all-text webpages (NSS = 1.97), followed by the pictorial (NSS = 1.77) and mixed (NSS = 1.80) webpages. 
\rev{While the difference in NSS scores was not significant across webpage types for the similarity between BubbleView clicks and eye fixations, the NSS scores were consistently higher for the text webpages. Only for one case, with a bubble size of 30 pixels and 10 seconds of viewing, were the NSS scores for the pictorial webpages the highest (Supplemental Material). This provides evidence that clicks tend to be more consistent with fixations on text elements.}


\begin{myframe}
\textit{Take-aways:} Both fixation and click data is more varied on webpages. Webpage images with lower IOC scores (more eye tracking variability) also had worse BubbleView similarity scores. Normalizing for IOC, BubbleView clicks can account for 78\% of eye fixations on webpages (as for natural images).
\end{myframe}

\rev{
\noindent \textit{Results on time, bubble size, and task}\\
We ran a two-way ANOVA (time $\times$ bubble size) on Exp. 3.1. The main effect of time on the CC and NSS scores was significant [CC: F(1,300)=19.25, $p<.01$, NSS: F(1,300)=9.65, $p<.01$], respectively but the effect of bubble size was not [CC: F(2,300)=1.92, n.s., NSS: F(2,300)=1.14, n.s.]. The interaction effect between time and bubble size was significant [CC: F(2,300)=6.95, $p<.01$, NSS: F(2,300)=3.35, $p<.05$].}

\rev{With a viewing time of 10 seconds, a bubble size of 30 pixels was too small, achieving significantly lower CC scores than bubble sizes of 50-70 pixels ($p<.05$). With a viewing time of 30 seconds, however, a bubble size of 70 pixels was too large, achieving significantly lower CC scores than bubble sizes 30-50 pixels ($p<.01$). No significant differences were found among the NSS scores (Table~\ref{tab:fiwi}).}
There exists a trade-off: with a longer viewing time, a smaller bubble radius provides more consistent clicks among participants; when limited by a shorter time, a larger bubble size becomes necessary. 

\rev{Given a bubble size of 30-50 pixels, the CC scores were significantly higher for a task duration of 30 seconds compared to 10 seconds ($p<.05$). The difference in NSS scores was only significant for the bubble size of 30 pixels. No significant differences were found with a bubble size of 70 pixels.}
Overall, BubbleView click maps generated with longer task durations of 30 seconds or longer (including with a description task) better approximated eye fixations than with a 10 second task duration. From this we conclude that information-dense images like websites require either longer viewing times or better defined tasks than free-viewing.

From Exp. 3.2, we found that for small numbers of participants ($n < 12$), the description task generated BubbleView click maps more similar to ground-truth eye fixations than the free-viewing task under all settings (Figure~\ref{fig:websaly_graphs}). The difference between the tasks is larger for smaller number of participants, and decreases with each extra participant. The click data tends to converge faster when a targeted task like description is used. \rev{However, this advantage disappears with more participants and a longer task time (30 sec, 30 pixel bubble radius). A description task takes longer and is more expensive to run, but might be a better choice when few participants are available.}

\begin{myframe}
\textit{Take-aways:} the less viewing time available, the larger the bubble size should be in order to better approximate free-viewing fixations. For a study with fewer participants, a description task is better than a free-viewing task. 
\end{myframe}

\begin{table}[]
\centering
\tbl{We evaluated BubbleView clicks at approximating ground-truth eye fixations on the FiWI dataset. BubbleView maps were computed with 12 participants for all experiments below. The score of the BubbleView maps predicting the ground-truth fixation maps is reported in CC, and the score of the BubbleView maps predicting the discrete fixation locations is reported in NSS. Normalized NSS is calculated by normalizing the NSS score by the inter-observer consistency (IOC) of the eye tracking participants. }{
\begin{tabu}to 1.0\textwidth{X[2.0,c] X[0.5,c] X[1.0,c] X[0.5,c] X[0.5,c]  X[1.0,c]}
\toprule
\multicolumn{1}{c}{\begin{tabular}[c]{@{}c@{}}Exp. 3: webpages\\ (ground-truth IOC: 1.85)\end{tabular}} & \multicolumn{1}{c}{\begin{tabular}[c]{@{}c@{}}Time \\ (sec)\end{tabular}} & \multicolumn{1}{c}{\begin{tabular}[c]{@{}c@{}}Bubble Radius\\ (pixel)\end{tabular}} & \multicolumn{1}{c}{CC} & \multicolumn{1}{c}{NSS} & \multicolumn{1}{c}{\begin{tabular}[c]{@{}c@{}}Normalized\\ NSS\end{tabular}} \\ \midrule\midrule
Free-viewing & 10 & 30 & 0.52 & 1.20 & 65\% \\\midrule
Free-viewing & 10 & 50 & 0.57 & \begin{tabular}[c]{@{}c@{}}1.34\\ \end{tabular} & 72\% \\\midrule
Free-viewing & 10 & 70 & 0.56 & 1.30 & 70\% \\\midrule
Free-viewing & 30 & 30 & 0.63 & 1.45 & 78\% \\\midrule
Free-viewing & 30 & 50 & 0.61 & \begin{tabular}[c]{@{}c@{}}1.41 \\ \end{tabular} & \begin{tabular}[c]{@{}l@{}}76\%\\\end{tabular} \\\midrule
Free-viewing & 30 & 70 & 0.57 & 1.32& 71\% \\\midrule
Description & unlim. & 30 & 0.63 & \begin{tabular}[c]{@{}c@{}}1.46\\ \end{tabular} & \begin{tabular}[c]{@{}l@{}}79\%\\\end{tabular} \\ \bottomrule
\end{tabu}
}
\label{tab:fiwi}
\end{table}

\section{Experiments comparing BubbleView to related methodologies}
\subsection{Experiment 4: comparison to importance annotations on graphic designs}\label{sec:exp4}

We hypothesized that the regions on an image where participants click using the BubbleView methodology correspond to the most important regions of the image. To test this hypothesis, we used the GDI dataset~\cite{odonovan} which comes with explicit importance annotations, where participants were instructed to annotate the image regions they considered important in graphic designs. We used this dataset to evaluate whether the number of BubbleView clicks on image regions corresponds to explicit judgements of importance. 

\begin{myframe}
\noindent \textit{Motivating questions}\vspace{-0.5em}
\begin{itemize}[label=\ding{212}]
\item Does BubbleView generalize to graphic designs?
\item Do BubbleView clicks correspond to regions of importance on graphic designs?
\end{itemize}
\end{myframe}

\noindent \textit{Stimuli} \\
The Graphic Design Importance (GDI) dataset contains 1,075 single-page graphic designs (e.g., advertisements, flyers, and posters consisting of text and graphical elements), collected from Flickr~\cite{odonovan}. 
No eye movements were collected for this dataset. \citeN{odonovan} highlighted two downsides of eye movements for this type of data: (1) fixations vary significantly over individual elements (like text blocks) even though those regions should have a uniform importance, and (2) eye fixations may occur in unimportant regions as a design is scanned and do not reflect conscious decisions of importance. Instead, 35 MTurk participants were asked to label important regions with binary masks, and these masks were averaged over all participants to produce a final importance map per design. \citeN{odonovan} noted that although importance maps produced by individual users are noisy, the average map gives a plausible relative ranking over design elements. 

\begin{figure}
 \centering
\includegraphics[width=1\textwidth]{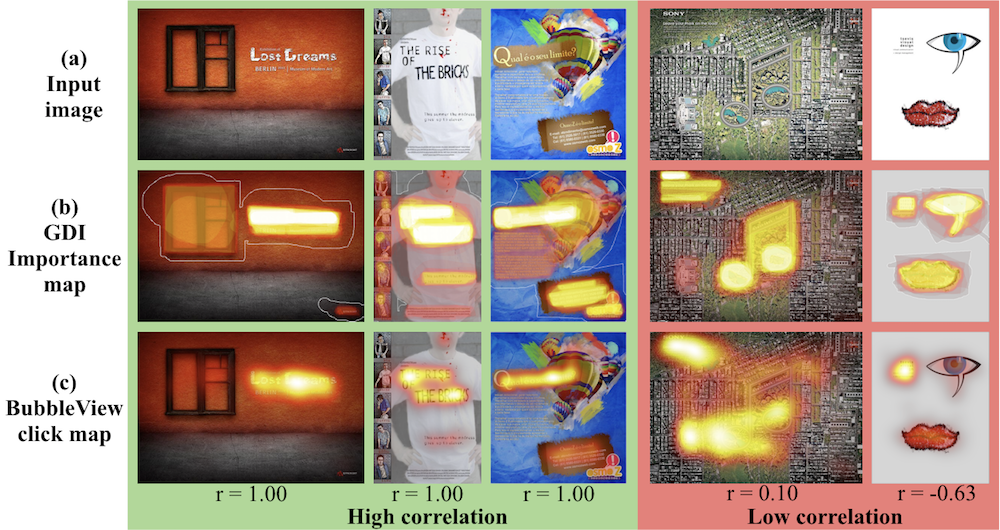}
 \caption{Example images from the GDI dataset. Images from the dataset (a), along with the provided explicit importance annotations (b). We show cases where BubbleView maps have high correlation, and cases with low correlation, to the importance annotations, in terms of how design elements are ordered by importance (c).}
 \label{fig:datasets4}
\end{figure}

We sampled 51 images from the GDI dataset at the original resolution of 600 $\times$ 400 pixels (Figure~\ref{fig:datasets4}). We blurred the images with a sigma of 30 pixels, manually chosen to distort text beyond recognition. \\ 

\noindent \textit{Method} \\
We ran an experiment with a bubble radius of 50 pixels and viewing time of 10 seconds, in which participants were asked to free-view each graphic design. BubbleView strikes a balance between eye fixations and explicit importance judgements for these images: (1) like fixations, clicks are collected in a free-viewing setting and are not uniform over design elements, but (2) like explicit annotations, the decisions of where to click reflect conscious decisions of importance. We collected an average of 15 participants worth of BubbleView click data for each image. \\

\noindent \textit{Analysis} \\
Unlike the quantitative evaluations in the previous sections, we did not directly compare the BubbleView click maps to the graphic design importance (GDI) maps. The spatial distributions of the explicit importance annotations in the GDI dataset are different from the click maps generated by our methodology. By construction, the importance annotations are uniform over design elements in the GDI dataset, while BubbleView clicks are not. For a fairer comparison, we computed the importance values each methodology assigns to different elements within each design (similar to the analysis at the end of Section~\ref{sec:exp1}).

\begin{figure}[h]
 \centering
\includegraphics[width=0.9\textwidth]{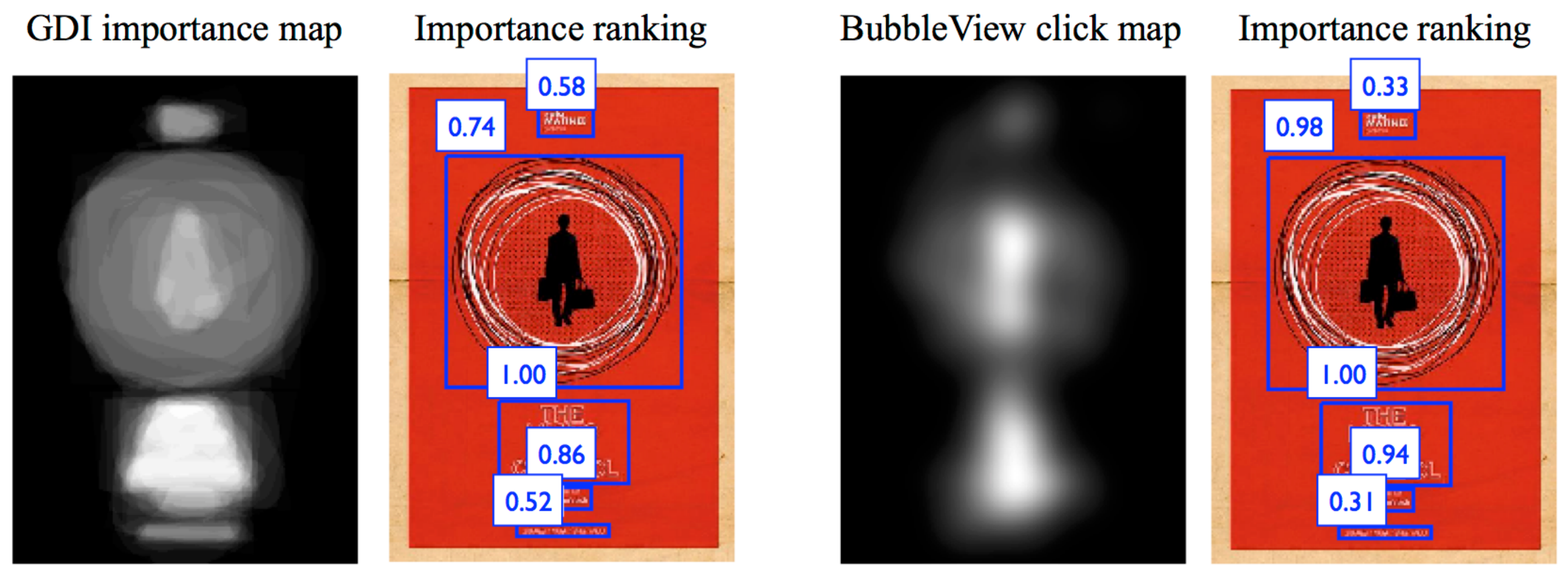}
 \caption{Importance maps were overlapped with element bounding boxes (outlined in blue) and the maximum map value per box was taken to be the importance score for that element (scores are the numbers above each box). Maps were first normalized to have values between 0 and 1, so the importance scores for all the graphic design elements also fall within the same range, where 1 corresponds to the most important element. In the case of the GDI importance map, MTurk workers made explicit judgements about aspects of the graphic design they considered the most important. A region of a graphic design has an importance score of 1 if all MTurk workers labeled that element as important. In the BubbleView study, MTurk workers clicked a blurred graphic design to expose small regions of the design at full resolution. A region of a graphic design has an importance score of 1 if the density of MTurk clicks in that region was highest.
}
 \label{fig:gdi-vs-bubbleview}
\end{figure}

We used bounding boxes to manually annotate all the elements in the 51 graphic designs chosen. For each design we normalized the GDI ground-truth importance map and the BubbleView click map. We took the maximum value of each map within an element's bounding box as the importance score of that element~\cite{jiang2015salicon,bylinskii2016should}. We correlated the importance scores assigned by both methodologies to the elements in each design (Figure~\ref{fig:gdi-vs-bubbleview}). \\


\ignore{

}

\noindent \textit{Results} \\
Across all 51 graphic designs, we achieved an average Pearson correlation of 0.66 and an average Spearman (rank) correlation of 0.60 between the element importance scores as assigned by BubbleView versus the original GDI annotations. Over 70\% of graphic designs had a correlation over 0.4. BubbleView importance maps can reasonably approximate explicit importance judgements for ranking elements of graphic designs, although there are some differences. For instance, the blurring of the image may interfere with visual features seen at different scales, as in the last two example images in Figure~\ref{fig:datasets4}. Depending on the blur, certain visual elements might not be clicked on (e.g. in Figure~\ref{fig:datasets4}, the \emph{note} because it blended into the background when blurred; the \emph{eye} because it was already visible in the blurred version). \\

\begin{myframe}
\textit{Take-aways:} BubbleView can be used to rank graphic design elements by importance. However, due to the varied feature sizes, blurring might significantly impact which design regions are clicked. 
\end{myframe}

\subsection{Experiment 5: comparison to mouse movements on natural images}\label{sec:exp5}

The most similar methodology to BubbleView is SALICON~\cite{jiang2015salicon}, which was introduced at roughly the same time\footnote{The SALICON and BubbleView methodologies were introduced a few months apart, but to different communities: \citeN{jiang2015salicon} to computer vision and \citeN{kim2015crowdsourced} to human-computer interaction.}. SALICON is also intended to be used in a crowdsourcing setting to approximate eye fixations~\cite{jiang2015salicon}. The differences are that SALICON captures continuous mouse movements, instead of clicks, and images are blurred adaptively, with a multi-resolution blur, recomputed for each cursor position. We investigated whether BubbleView click maps are similar to SALICON mouse movement maps, when averaged over multiple participants.
Because the SALICON blur is multi-resolution and adaptive, we experimented with different blur sigmas and bubble sizes in BubbleView, to find a fixed setting of parameters that best approximates the SALICON viewing conditions. We also compared SALICON and BubbleView at approximating eye fixations collected in a controlled lab setting, since both methodologies are presented as alternatives to eye tracking. 
\begin{myframe}
\noindent \textit{Motivating questions}\vspace{-0.5em}
\begin{itemize}[label=\ding{212}]
\item Under what settings does BubbleView most closely match SALICON?
\item Which methodology better approximates eye fixations on natural images?
\end{itemize}
\end{myframe}

\begin{figure}
 \centering
\includegraphics[width=1\textwidth]{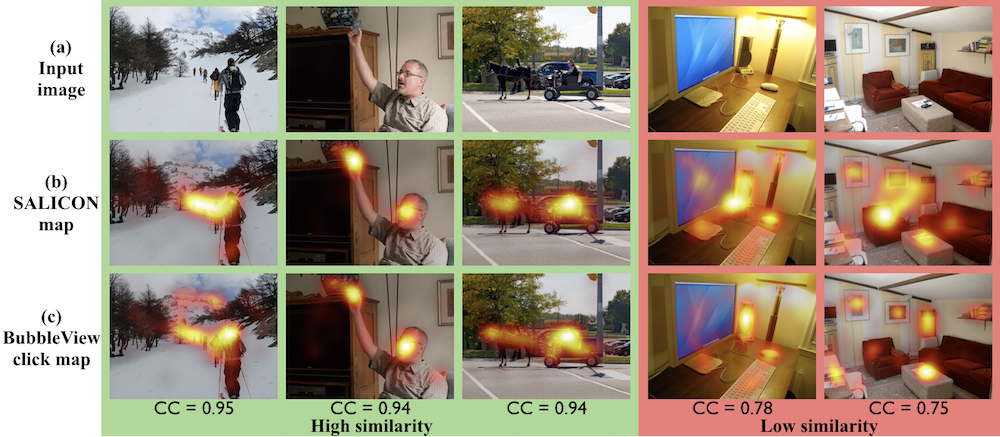}
 \caption{Example images from the SALICON dataset. Example dataset images (a), and ground truth mouse movements collected by SALICON (b). We show cases where BubbleView maps have high similarity, and cases with low similarity, to SALICON maps (c).}
 \label{fig:datasets5}
\end{figure}

\noindent \textit{Stimuli} \\
The SALICON dataset consists of mouse movements collected on 20K MS COCO (Microsoft Common Objects in Context) natural images \cite{lin2014microsoft}. In the original study, mouse movements were collected on Amazon's Mechanical Turk by presenting images to participants for 5 seconds each and allowing them to freely explore each image by moving the mouse cursor. 
We randomly sampled 51 images at the original image size of 640 $\times$ 480 pixels from the SALICON dataset (Figure~\ref{fig:datasets5}).
\\ 

\begin{figure}
 \centering
\includegraphics[width=1\textwidth]{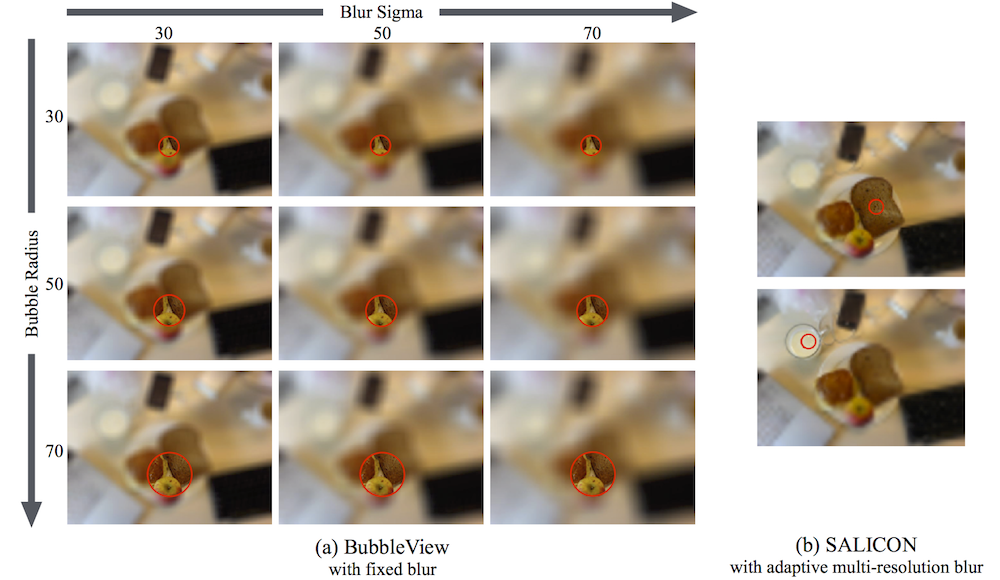}
 \caption{We used 9 different parameter settings in our BubbleView experiments, on images from the SALICON dataset (a). We wanted to find a fixed setting of bubble size and blur to mimic the adaptive multi-resolution blur used in the SALICON methodology (b). The rightmost figure is from~\protect\citeN{jiang2015salicon}.}
 \label{fig:experiment5-setting}
\end{figure}

\noindent \textit{Method} \\ 
In \textbf{Exp. 5.1}, we used a $3\times3$ factorial design (blur sigma: 30, 50, and 70 pixels; bubble radius: 30, 50, and 70 pixels; see Figure~\ref{fig:experiment5-setting}). Using a free-viewing task, we had participants explore each image for 10 seconds each. We wanted to account for longer times to click, rather than move, the mouse. 

To disentangle the influence of mouse clicks/movements versus fixed/adaptive blur on the methodology differences between SALICON and BubbleView, we ran \textbf{Exp.~5.2}, using BubbleView with a moving-window approach like SALICON, but maintaining a fixed blur kernel. In this setup participants used mouse movements to reveal image regions at normal resolution. We had two experiment conditions (bubble radius sizes of 30 and 50 pixels) with a fixed blur sigma of 30 pixels (found appropriate in Exp. 5.1) and viewing time of 5 seconds (as in SALICON). 
We collected an average of 15 participants worth of BubbleView click data for each image under each condition.\\

\begin{table}[]
\centering
\tbl{
\rev{We evaluated BubbleView click maps (with $n=12$ participants per image) at approximating SALICON mouse movements, measured using CC and NSS metrics. Normalized NSS is computed by taking into account the IOC of the SALICON participants (NSS = 1.50). Both bubble radius and blur sigma are measured in pixels. BubbleView with a blur radius of 70 pixels achieved significantly lower CC scores than with other blur settings ($p<.01$ for all bubble sizes). The other differences were not significant.}
}{
\begin{tabu}to 1.0\textwidth{X[1.0,c] X[1.0,c] X[1.0,c] X[1.0,c] X[1.0,c] X[1.0,c]}
\toprule
 &  &  & \multicolumn{3}{c}{Blur Sigma (pixel)} \\ \midrule
 &  &  & 30 & 50 & 70 \\ \midrule \midrule
\multirow{3}{*}{\begin{tabular}[c]{@{}c@{}}Bubble radius \\ (pixel)\end{tabular}} & 30 & \begin{tabular}[c]{@{}r@{}}CC\\ NSS\\ Normalized NSS\end{tabular} & \begin{tabular}[c]{@{}c@{}}0.84\\ 1.21\\ 81\%\end{tabular} & \begin{tabular}[c]{@{}c@{}}0.84\\ 1.15\\ 77\%\end{tabular} & \begin{tabular}[c]{@{}c@{}}0.78\\ 1.06\\ 71\%\end{tabular} \\ \cmidrule(l){2-6} 
 & 50 & \begin{tabular}[c]{@{}r@{}}CC\\ NSS\\ Normalized NSS\end{tabular} & \begin{tabular}[c]{@{}c@{}}0.86\\ 1.23\\ 82\%\end{tabular} & \begin{tabular}[c]{@{}c@{}}0.84\\ 1.15\\ 77\%\end{tabular} & \begin{tabular}[c]{@{}c@{}}0.80\\ 1.04\\ 69\%\end{tabular} \\ \cmidrule(l){2-6} 
 & 70 & \begin{tabular}[c]{@{}r@{}}CC\\ NSS\\ Normalized NSS\end{tabular} & \begin{tabular}[c]{@{}c@{}}0.84\\ 1.20\\ 80\%\end{tabular} & \begin{tabular}[c]{@{}c@{}}0.84\\ 1.11\\ 74\%\end{tabular} & \begin{tabular}[c]{@{}c@{}}0.79\\ 1.04\\ 69\%\end{tabular} \\ \bottomrule

\end{tabu}
}
\label{tab:move}
\end{table}

\noindent \textit{Results on using BubbleView to approximate SALICON} \\
\rev{We ran a two-way ANOVA (blur $\times$ bubble size) on Exp. 5.1. The main effect of bubble size was not significant [CC: F(2,450)=2.28, NSS: F(2,450)=0.19, n.s.] (as found in Exp. 1 and 3.1). The main effect of blur on scores was significant [CC: F(2, 450)=19.97, $p<.01$, NSS: F(2,450)=6.86, $p<.05$]. BubbleView with a blur radius of 70 pixels achieved significantly lower CC scores than with other blur settings ($p<.01$ for all bubble sizes).
We did not find an interaction effect between blur and bubble size [CC: F(4,450)=0.44, NSS: F(4,450)=0.04, n.s.].}
We found highest similarity between BubbleView click maps and SALICON maps at bubble radius sizes of 30--50 pixels and blur sigma of 30--50 pixels (Table \ref{tab:move}), for which the normalized NSS scores ranged from 77\% to 82\%. 

\begin{figure}[h]
 \centering
\includegraphics[width=0.8\textwidth]{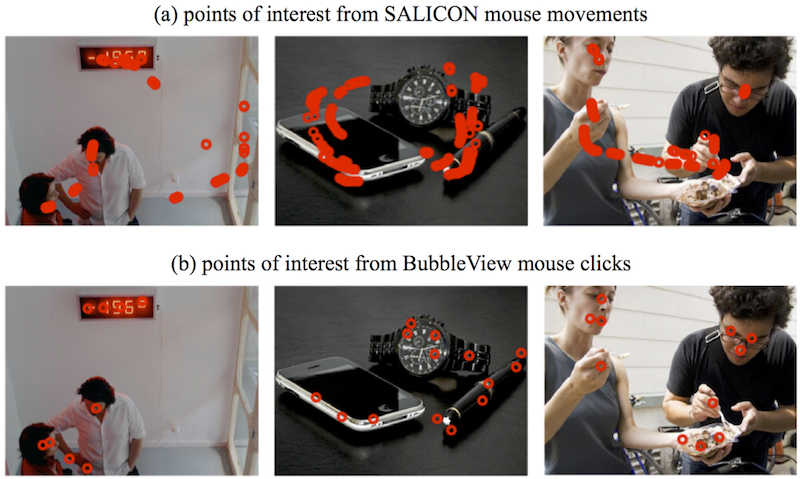}
 \caption{When participants can move the mouse anywhere on the image without having to click, the collected data contains motion traces as byproducts (a). Instead of only capturing the points of interest in an image where an observer's attention stops, the moving-window approach also captures the transitions between these regions, which are less relevant and add noise to the data. Although these trajectories can be post-processed into discrete regions of interest, our approach is to directly collect participant mouse clicks on points of interest, with no further post-processing required (b).}
 \label{fig:bubble-movement}
\end{figure}

What are the remaining differences? Using mouse movements, more points of interest are generated than using clicks. Many of the points sampled using mouse movements occur in the transition between regions in an image, and might be introducing noise into the data (Figure~\ref{fig:bubble-movement}). This suggests that a different threshold might be more effective at converting continuous mouse movements into discrete points of interest. An advantage of the BubbleView clicks is that no such post-processing is necessary, since the clicks directly correspond to points of interest.

In Exp. 5.2, we modified BubbleView to collect continuous mouse movements and shortened the time per image to 5 sec, such that the only remaining difference with SALICON was the treatment of blur. \rev{We observed that the mean number of samples was 143.02 (SD=13.14) using the sampling rate of 100 Hz, which translates to 14,302 raw samples, on average, per participant. This is significantly larger than the mean click count of 13.09 (SD=1.38) per participant in Exp 5.1.}

With the moving-window BubbleView setting the scores were: for bubble size 30: CC: 0.87, NSS: 1.21, normalized NSS: 81\%; bubble size 50: CC: 0.88, NSS: 1.24, normalized NSS: 83\%. Compared to the clicks, these scores were not statistically significantly different [$F(200)<2.2$, n.s.].
In other words, BubbleView can approximate SALICON with or without mouse movements.  
Importantly, BubbleView can approximate SALICON without requiring a multi-resolution adaptive blur, simply with a single fixed blur setting. Our fixed blur setting is much less computationally expensive and does not require the pre-study system checks as in~\citeN{jiang2015salicon}. 

\begin{myframe}
\textit{Take-aways:} BubbleView with a bubble size of 30--50 pixels and a blur sigma of 30--50 pixels can approximate the continuous mouse movements and adaptive, multi-resolution blur of the SALICON methodology. 
\end{myframe}

\begin{figure}
 \centering
\includegraphics[width=0.6\textwidth]{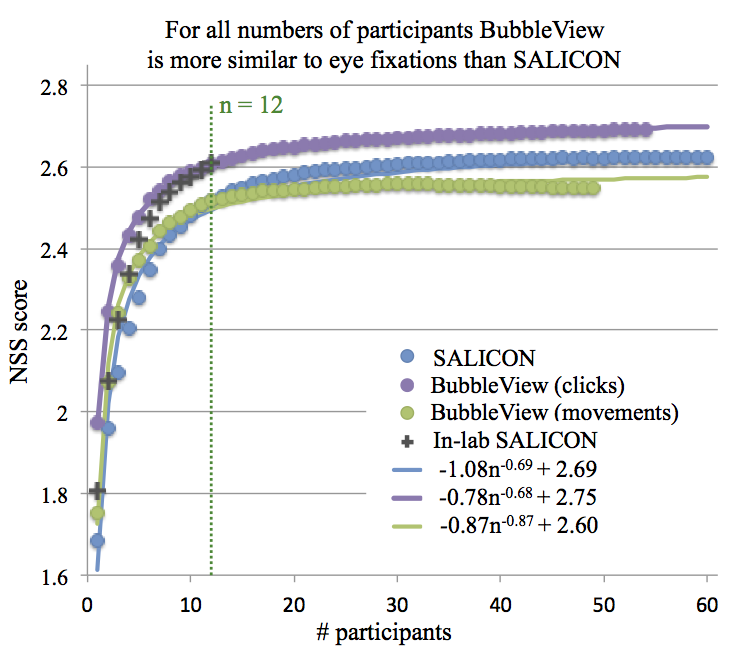}
 \caption{The NSS score obtained by comparing mouse clicks and mouse movements to ground truth eye fixations on natural images in the OSIE dataset. We compare mouse clicks gathered using BubbleView on MTurk (purple), mouse movements gathered using BubbleView on MTurk (green), mouse movements gathered using SALICON on MTurk (blue), and mouse movements gathered using SALICON in a controlled lab setting (black crosses). Each point represents the score obtained at a given number of participants, averaged over \rev{10 random splits of participants and} all 51 images used. 
 }
 \label{fig:osie_extrap}
\end{figure}

\noindent \textit{Results on using both methodologies to approximate eye fixations} \\
In Exp. 2.2 we compared BubbleView clicks and mouse movements to SALICON mouse movements at approximating ground truth eye fixations on 51 OSIE images. The BubbleView click maps (with $n=12$ participants, bubble radius of 30 pixels) achieve NSS = 2.61 (CC = 0.81) at predicting ground-truth fixation maps, compared to SALICON mouse movement maps which achieve NSS = 2.52 (CC = 0.81). \rev{It takes over 30 SALICON participants to achieve the same similarity to fixation maps as 12 BubbleView participants (Figure~\ref{fig:osie_extrap}). 
Replacing BubbleView clicks with mouse movements actually decreases performance: NSS = 2.52 (CC = 0.81), but this drop in performance is not significant at the $p=.05$ level}. \rev{For all feasible numbers of participants ($n < 60$ in Figure~\ref{fig:osie_extrap}), BubbleView offers a better approximation to eye fixations than SALICON.}

Data was also available for 12 in-lab participants who used the SALICON methodology to view images in a controlled lab setting~\cite{jiang2015salicon}. The in-lab SALICON maps, which capture these mouse movements, achieve NSS = 2.61 (CC = 0.81) when compared to fixation maps, the same score as our BubbleView maps (Table~\ref{tab:osie}). 
\rev{From Figure~\ref{fig:osie_extrap} we can see that the performance of the in-lab SALICON is increasing at a greater rate than either BubbleView or online SALICON. However, more in-lab SALICON participants would be needed to see whether this trend continues. In any case, it requires a controlled lab setting, which we aim to avoid.}


\begin{myframe}
\textit{Take-aways:} On a natural image dataset, BubbleView clicks better approximate eye fixations than SALICON mouse movements for all feasible numbers of participants ($n<60$). BubbleView performed better with clicks than BubbleView with mouse movements.
\end{myframe}

\section{Discussion}
\label{sec:discussion}


\textbf{Similarity of BubbleView clicks to eye fixations:} We
showed that across 3 different image types (information visualizations, natural images, and static webpages) and 2 types of tasks (free-viewing and description), BubbleView clicks provide a reasonable approximation to eye fixations collected in a controlled lab setting. Specifically, across all these image types BubbleView clicks accounted for over 75\% of eye fixations when only 10--15 BubbleView participants were used (Tables II-IV). Of all settings, BubbleView clicks provided the best approximation to eye fixations on information visualizations with a description task, accounting for up to 90\% of eye fixations with only 10 participants, and 92\% with 20 participants (Table II). On both natural images and websites, BubbleView clicks could account for up to 78\% of eye fixations with 10--12 participants (Tables III, IV). The fixations of eye tracking participants were much more consistent on the natural images than on the websites, so the viewing behavior on natural images should be easier to predict. \rev{Despite the remaining gap between BubbleView clicks and eye fixations for natural images and webpages, the fact that already 10--15 BubbleView participants achieves a reasonable approximation to fixations is promising for perception studies that might otherwise require specialized eye-tracking hardware.}

\begin{figure}
 \centering
\includegraphics[width=1\textwidth]{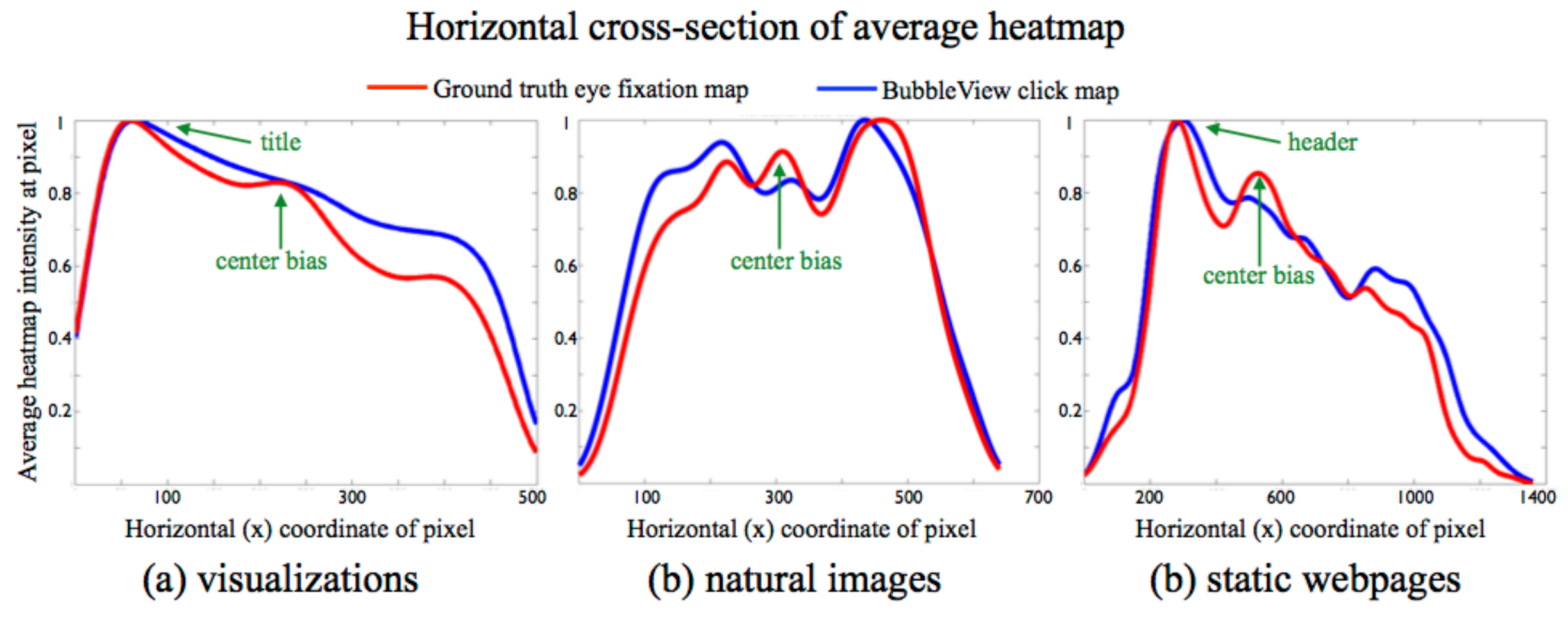}
 \caption{Taking a horizontal cross-section of the average BubbleView click map and the average fixation map across 51 images on 3 datasets, we see the fixation map has a consistent center bias. This replicates the analysis used by \protect\citeN{Tatler2007central} to report on human fixation bias in natural images. This bias emerges as a peak near the center of an image, which corresponds to the midway point along the $x$-axis in each of these plots. The BubbleView click map does not have this bias, which accounts for some of the systematic differences observed between the click and fixation maps. At the same time, the bubble clicks tend to capture the same general characteristics as fixations, for instance of increased attention in the leftmost parts of visualizations and webpages, corresponding to the titles and headers. }
 \label{fig:spatialbias}
\end{figure}

\textbf{Remaining differences between BubbleView clicks and eye fixations:} Part of the remaining gap between BubbleView clicks and eye fixations is that BubbleView does not capture the unconscious movements of the eyes due to bottom-up, pop-out effects, or systematic biases. One such systematic bias commonly referred to in the eye tracking literature is center bias \cite{Tatler2007central,Borji_2012,bylinskii2015towards}, whereby a relatively high number of fixations occur near the center of the image. One explanation for such bias is that it is part of an optimal viewing strategy that is involved in planning successive fixations. By averaging fixation maps across dataset images, we can see a peak near the spatial center of the image emerge across the eye fixations, but not the BubbleView clicks (Figure~\ref{fig:spatialbias}). Because BubbleView naturally slows down the exploration task by making participants consciously decide where to click next, it captures higher-level viewing behaviors not as affected by systematic biases. We recommend using BubbleView with a well-defined task, like describing the content of the visual input, to measure which regions of that visual input are most important or relevant for the task. 

\rev{A recent paper by~\citeN{tavakoli2017saliency} analyzes some of the semantic differences between eye fixations and mouse movements on the OSIE dataset, by taking into account annotated image regions. They find that there tends to be more disagreement between eye fixations and mouse movements in background regions of the image.} 


\textbf{Effect of BubbleView parameters:} The BubbleView click maps were quite robust across different parameter settings. We did not find significant effects of bubble radius on the resulting BubbleView clicks (Exp. 1,3,5). 
Across all our experiments (Exp. 1--5), we found that a blur kernel sigma in the range of 30--50 pixels was appropriate for all of our image types, where we manually selected a sigma value for each image dataset to ensure that text was unintelligible when blurred and would require explicit clicking on to read. In other words, to mimic peripheral vision, the blur level was chosen to eliminate legible details beyond the focal region. \rev{However, a blur sigma with a 70-pixel radius was too high, and seemed to hinder exploration of the image by eliminating too much context, as similarity of BubbleView clicks to eye fixations significantly dropped for this blur level compared to a blur of 30--50 pixels (Exp. 5).}

We found that a bubble radius in the range of 30 to 50 pixels seems to consistently work best for different image types and image sizes that comfortably fit within the browser window (ranging from 500 $\times$ 500 to 1000 $\times$ 600 pixels). Here ``best'' refers to the ability of BubbleView clicks to most closely approximate fixations on images with the smallest number of participants. Smaller bubble sizes lengthened the duration and effort for completing the task, for the same quantitative results. 
Our chosen bubble sizes typically corresponded to 1--2 degrees of visual angle as measured in the corresponding eye tracking experiments. A bubble size of 1--2 degrees of visual angle mimics the size of the foveal region during natural viewing.

However, bubble radius is also intricately related to task timing and image complexity (Exp. 3). The more content there is on an image to look at, the more time that is required; the smaller the bubble, the more clicks to explore all of the content. A larger bubble radius can compensate for less available time, because each click exposes more of the image. For best results, we recommend a smaller bubble radius but longer task time. In our studies, the longest time for free-viewing tasks was 30 seconds (Exp. 3). For description tasks, participants spent an average of 1.5--3 minutes per image, clicking and describing (Exp. 1,3).

\rev{The number of clicks participants made decreased with increasing bubble size, even though the time for the task stayed the same. We observed this trend across all of our experiments. On average, 1--1.5 clicks were made per second in the BubbleView setup, compared to an average of 2--3 fixations per second in eye tracking studies. The BubbleView setup (when implemented with clicks) slows down visual processing so about half as many interest points are examined every second.} 


The best prediction performance overall occurs in the setting of a well-defined task, such as describing the visual content of an image. However, tasks must be well-matched to the images used. For instance, asking participants to describe an information visualization is well-defined because each of the visualizations we used had a main message that was being communicated (Exp. 1). On the other hand, we did not use the description task for the graphic designs (Exp. 4), because it was harder to objectively define what should be described. 

\rev{\textbf{Number of participants, task, and data quality:} For our tasks, we found 10--15 participants sufficient, accounting for over 97\% of the performance achievable with 40 participants (Exp. 1,2), where performance is measured by how many of the eye fixations can be approximated by clicks. The more participants, the better the data, as noisy clicks get averaged out. However, when there is a constraint on how many participants can be recruited/afforded, a more involved task (like asking the participant to provide a text description) can result in cleaner data (Exp. 3). Such a task adds an energy barrier to clicking: to minimize effort, participants are more likely to click on image regions informative for completing the task, rather than randomly.
}




\textbf{Mouse clicks versus movements: } We compared our methodology of collecting discrete mouse clicks to SALICON's moving-window approach~\cite{jiang2015salicon} in Exp.~5. 
\rev{We found that for any number of participants less than 60}, BubbleView is a better approximation to ground truth eye fixations (Figure~\ref{fig:osie_extrap}). \rev{This is similar to the task, data-quality trade-off discussed above. Clicks add an energy barrier to action: since clicking takes more effort than moving the mouse, participants are more selective about where they click. As a result,}
BubbleView provides cleaner data with fewer artifacts, such as the byproducts of continuous mouse movements (Figure~\ref{fig:bubble-movement}). Furthermore, the moving-window methodology requires post-processing to differentiate mouse positions corresponding to points of interest from transitions. Collecting clicks directly eliminates such post-processing steps.


On the other hand, \rev{a byproduct of the} higher effort of clicking on an image area rather than moving a mouse over it, is that fewer image areas will be explored by clicking. 
If the focus of the study is to select the most important regions in an image, then clicks should suffice. 
In Table~\ref{tab:tradeoffs} we summarize the tradeoffs between the two methodologies. We note additionally that we were able to approximate SALICON's multi-resolution adaptive blur with a single, fixed blur (Exp. 2,5) to achieve similar performances at much lower computational cost.

\begin{table}[h]
\centering
\tbl{Comparison of BubbleView and SALICON~\protect\cite{jiang2015salicon}. SALICON consists of capturing continuous mouse movements on an image with adaptive multi-resolution blur. The blur is continuously recomputed for every mouse location at 100 Hz. Continuous mouse tracks are discretized into points of interest using experimenter-specified thresholds. In BubbleView, discrete mouse clicks are collected on an image with a fixed blur. This is easier to implement and has fewer computational limitations. No additional post-processing is required. The collected BubbleView data is less noisy and converges faster, although clicking takes more time.}{

\begin{tabu}to 1.0\textwidth{X[1.4, c] X[0.8, c] X[0.8, c]}
\toprule
\textbf{Property} & \textbf{BubbleView} & \textbf{SALICON} \\ \midrule \midrule
Speed of convergence to eye fixations & faster & slower \\
Number of participants required & fewer & more \\
Time per task & higher & lower \\
Post-processing & less & more \\
Computational cost & less & more \\ \bottomrule
\end{tabu}
}
\label{tab:tradeoffs}
\end{table}

\textbf{BubbleView for image importance:} The density of clicks in different image regions roughly corresponds to the importance of those regions. Specifically, across a collection of graphic designs, BubbleView clicks on different design elements correlated with explicit importance judgements made on the same designs (Exp. 4). BubbleView clicks ranked visualization elements similarly to eye fixations (Exp. 1). Thus, BubbleView can be used not only to derive conclusions about human perception (where people look), but also to make general conclusions about images and designs: how is importance distributed across an image? Which design elements are most important? This knowledge can in turn can be leveraged for design applications~\cite{predimportance}.

\textbf{Data quality and filtering: } BubbleView participants were quite consistent with each other in where they clicked, leading to a relatively fast convergence of the aggregate BubbleView click maps to ground truth eye fixation maps. For most of our experiments, we found about 10-15 participants provided enough click data to reasonably approximate eye fixations. 

After collecting the BubbleView data, we performed a number of filtering steps, including throwing out participants who did not click a minimum number of times and additional clicking outliers. This filtering of participants and bubbles lead to a data reduction of only 2\% on average, indicating that initial data quality was pretty high (Supplemental Material). 

The description task has the additional benefit of providing another filtering layer: if a participant-provided description is evaluated as poor, we can assume that they did not do the task with sufficient thoroughness, or clicked in regions of the image that were irrelevant for the task. This filtering step can either be performed manually by the experimenter or implemented as a crowdsourcing task (e.g., by having Amazon Mechanical Turk workers rate descriptions by quality).

\textbf{Cost: } The price to obtain a BubbleView click map per image depends on the amount of time a participant spends on each image and the total number of participants recruited. The average hourly rate for Amazon's Mechanical Turk is \$6/hour, so we use \$0.1/min for our tasks. It is common to make MTurk tasks bite-sized (e.g., a few minutes to 10--15 min each)~\cite{kittur2008crowdsourcing}. Using these guidelines, we provide an approximate cost of obtaining a BubbleView click map per image using 10-15 participants. 
Table~\ref{tab:costs} contains a breakdown of costs that can be used as guidelines. 


\begin{table}[h]
\centering
\tbl{Total computed costs per image for obtaining the BubbleView clicks of 10--15 participants (both ends of the range included). These costs depend on how long, on average, participants spend on each image, which in turn depends on the task used. In the free-viewing setting, we fixed the time to either 10 or 30 seconds per image. In the description task, time is unconstrained, and participants move on to the next image after submitting their description for the previous image. During piloting, we estimated time per image for clicking and describing to take about 1.5 minutes. In reality, it took on average 3.2 minutes per image. The description task is more expensive but provides higher-quality click data and an additional data source: the descriptions themselves. These descriptions also serve as quality-control: the clicks of participants who generated poor-quality descriptions can be discarded. }{
\begin{tabu}to 1.0\textwidth{X[1, c] X[0.9, c] X[0.9, c] X[0.9, c] X[1, c] X[1, c]}
\toprule
\textbf{Task} & \textbf{Time/image} & \textbf{Images/HIT} & \textbf{Cost/HIT} & \textbf{Participants/HIT} & \textbf{Cost/Image} \\ \midrule \midrule
Free-viewing & 10 sec & 17 & \$0.30 & 10--15 & \$0.18--\$0.26 \\
Free-viewing & 30 sec & 17 & \$0.90 & 10--15 & \$0.53--\$0.79 \\
Description & 180 sec & 3 & \$0.50 & 10--15 & \$3.34--\$5.00 \\ \bottomrule
\end{tabu}
}
\label{tab:costs}
\end{table}

\textbf{Methodology limitations:} Compared to natural viewing or moving a mouse, clicking takes more time and effort, resulting in longer task timings and higher costs. Certain image regions which might not be as relevant to the task might never be clicked on, even though they may have received a quick glance in an eye tracking or moving-window setting. As a result, the image regions selected by clicks will tend to be more selective than the regions selected in these other settings. As shown in this paper, the advantage of this selectivity is cleaner, more consistent results across participants. This can be used for determining the most important regions in an image (Exp. 4). But this comes at the potential disadvantage of certain image regions being missed, and other regions, like text, receiving disproportionate clicks (Exp. 1,3). 
How to encourage a more diverse sampling of image regions while maintaining all the other advantages of BubbleView is a question for future investigations.


\section{Conclusion and future work}
\label{sec:conclusion}
In this paper we presented BubbleView, a mouse-contingent methodology to \rev{approximate eye fixations} using mouse clicks. We validated BubbleView by conducting a series of experiments on different image stimuli and comparing clicks to eye fixations, importance maps, and mouse movements. We showed that BubbleView can reasonably approximate fixations, be used to collect image importance driven by human perception, and has a number of advantages compared to the moving-window approach, including better performance with fewer participants. 

We analyzed BubbleView in the context of 4 image types (information visualizations, natural images, static webpages, and graphics designs), with 2 task types (free-viewing and description), with different task timing, image blur and bubble sizes, and different numbers of study participants. We provided the interested experimenter with some guidelines on how to use BubbleView for different tasks, how to select parameters, and which settings we found to work best under different conditions. Here we provide additional ideas of how BubbleView can be used and built on top of.

\textbf{Integrating BubbleView into crowdsourcing pipelines: } Unlike eye tracking experiments, BubbleView experiments can be feasibly ported online for the efficient and scalable collection of data using crowdsourcing. Large amounts of data call for data filtering and analysis methods that can scale as well.
As shown in this paper, BubbleView clicks can be analyzed automatically. In cases where text input is also collected from participants, filtering and analysis may require additional manual effort. However, it is possible to consider crowdsourcing pipelines where the data collected from the BubbleView tasks is piped directly into filtering tasks. 

Following the idea of question-answering tasks, BubbleView can be incorporated into multi-player crowdsourcing games (e.g., ESP Game \cite{von2004labeling}). For instance, one participant can generate questions, while the other participant answers using BubbleView clicks. In this setting, the first participant queries and supervises the responses of the second participant. In such a way both data collection and data cleaning can be built into the game.

\textbf{BubbleView data for training computational models: } BubbleView can be used to generate large datasets for training computational models. BubbleView click maps on images can be used as importance maps for those images, and computational models can learn from this data to make predictions for new images~\cite{predimportance}. While many saliency models have been developed for natural images, and many natural-image saliency datasets exist (Section~\ref{sec:saliency}), graphic designs and visualizations have not received as much attention. A saliency model based on these type of stimuli could open up many interesting applications such as extracting important information based on salient regions or providing design feedback~\cite{odonovan,o2015designscape,RosenholtzTAP,predimportance}.

\textbf{Measuring information content: } Clicking on an image region takes more effort than mousing over, and in turn, glancing at it. There is likely a relationship between the information content of an image region and the likelihood with which it is clicked, moused over, and glanced at. Clicking imposes a kind of energy barrier on the image content that will be explored by participants. Given a targeted task such as describing an image, participants are motivated to click in as few regions as necessary to reduce the overall effort and total task time. As a result, they tend to click in the most informative regions. Increasing the bubble size lowers this energy barrier: participants become less selective of where they're clicking when they can expose more of the image with each click. Changing the image blur also affects which image region will be clicked, based on its information content. More deeply studying the relationship between visual feature size, information content, image blur and bubble size is likely to provide some interesting insights. In the present study, by virtue of the images we selected for our experiments (e.g., to contain legible text) and the narrow range of image sizes we used, results were pretty stable across blur and bubble settings.

\textbf{Extending BubbleView to other tasks: }The interested experimenter may also choose to use BubbleView in settings and with parameters beyond the ones in this paper, which leaves many possibilities for future investigation. For instance, BubbleView can easily be extended to other visual attention tasks including visual search\footnote{Some examples of visual attention tasks with operational definitions and recommended evaluations are included in \citeN{bylinskii2015towards}.}. To implement a version of visual search using BubbleView, participants can be shown a blurred image and asked to find something in the image (e.g., an object in a natural scene, a specific piece of information in a graph, or an element in a graphic design). Task time can be either fixed, contingent on when the participant chooses to continue to the next image, or contingent on the participant's clicks (i.e., moving to the next image after the correct/expected location is clicked, or after a fixed number of clicks).

Another possible use for BubbleView is modifying the description task into a question-answering task. Participants can be asked to answer a specific question about the image by clicking around the blurred image to expose the content underneath. Each answer, correct, incorrect, or subjective, can be analyzed together with the sequence of clicks made (similar to \citeN{das2016human}).

While we originally designed BubbleView as a more efficient alternative to collecting eye fixations on images, we have also shown in this paper that it can be used to measure the importance of different image regions. This idea can be pushed even further in the future, using BubbleView to narrow in on image regions most useful for answering specific questions, extracting particular insights, or completing specific visual tasks. We showed that BubbleView generalizes to different types of images, including natural scenes, visualizations, websites, and graphic designs. This can be expanded to new image types, for instance for studying medical images, geographical maps, user interfaces, slides and posters. For future explorations, we provide our tool and code for launching experiments at \underline{massvis.mit.edu/bubbleview}.

\begin{acks}
The authors would like to thank Peter O'Donovan for sharing his dataset and for helpful discussions, and Ming Jiang for answering questions about the SALICON data and methodology. The authors would like to acknowledge Aaron Hertzmann, Bryan Russell, Jean-Daniel Fekete, and Lester Loschky for helpful input. \rev{Feedback of anonymous reviewers has helped to significantly improve the quality and clarity of the writing.}

This work has been made possible through support from Google, Xerox, the NSF Graduate Research Fellowship Program, the Natural Sciences and Engineering Research Council of Canada, and the Kwanjeong Educational Foundation. We also acknowledge the support of the Toyota Research Institute / MIT CSAIL Joint Research Center.

\end{acks}

\bibliographystyle{ACM-Reference-Format-Journals}
\bibliography{paper}

\end{document}